\documentclass[a4paper,11pt]{article}
\pdfoutput=1 

\usepackage{jheppub} 

\usepackage[T1]{fontenc} 

\usepackage{physics}
\usepackage{amsmath}
\usepackage{amsthm}
\usepackage{physics}
\usepackage{mathtools}
\usepackage{graphicx}
\usepackage{subcaption}
\usepackage{courier}
\usepackage{hyperref}
\usepackage{amsfonts}

\title{\boldmath Holographic Complexity in FRW Spacetimes}
\author{Reginald J. Caginalp}

\affiliation{Department of Physics, University of California, Berkeley, CA 94720, USA}

\emailAdd{caginalp@berkeley.edu}

\abstract{We examine the holographic complexity conjectures in the context of holographic theories of FRW spacetimes. Analyzing first the complexity-action conjecture for a flat FRW universe with one component, we find that the complexity grows as $t^2$, regardless of the value of $w$.  In addition, we examine the holographic complexity for a flat universe sourced by a scalar field that is undergoing a transition. We find that the complexity decreases when the holographic entanglement entropy decreases for this universe. Moreover, the calculations show that, while the entanglement entropy decreases only slightly, the magnitudes of the corresponding fractional decreases in complexity are much larger. This presumably reflects the fact that entanglement is computationally expensive. Interestingly, we find that the gravitational action behaves like a complexity, while the total action is negative, and is thus ill-suited as a measure of complexity, in contrast to the conjectures in AdS settings. Finally, the implications of the complexity-volume conjecture are examined. The results are qualitatively similar to the complexity-action conjecture. }

\begin{document} 
\maketitle
\flushbottom

\section{Introduction}
\label{sec:intro}

Recent work has revealed deep connections between between gravity, spacetime, and information. The AdS/CFT correspondence posits that any theory of quantum gravity in $d+1$-dimensional anti-de Sitter space (AdS) is equivalent to a conformal field theory (CFT) in $d$ dimensions \cite{Maldacena,Gubser,Witten}. This correspondence is a concrete realization of the holographic principle, which conjectures that the degrees of freedom in a theory of quantum gravity are encoded in one fewer dimension. 

The celebrated Ryu-Takayanagi formula \cite{RT} (and its covariant generalization, the Hubeny-Rangamani-Takayanagi formula \cite{HRT}) of the AdS/CFT correspondence posits an equivalence between entanglement entropy in a holographic CFT and minimal surfaces in the bulk. This has led to an improved understanding of the rich interplay between the spacetime in the bulk theory, and information in the boundary theory.

There has been much recent speculation about the possible role of complexity in the AdS/CFT correspondence. This arises from the following consideration. If we consider the maximum volume slice anchored at boundary time $t,$ of an AdS black hole, the volume will grow linearly with $t$. Moreover, it will (classically) grow forever. However, the entanglement entropy saturates after a relatively short time. Thus, there must be some CFT quantity that encodes this growth. Even after the saturation of the entanglement entropy, the quantum state continues to evolve subtly in time, and the entanglement entropy is too crude of a measure to detect these changes. A promising candidate for the CFT dual of the volume growth is the \textit{complexity}. Consider some state $\ket{\psi}$ in a Hilbert space $\mathcal{H}$, and some ``simple" reference state $\ket{\psi_0} \in \mathcal{H}$. For example, in a system of $n$ qubits, the state $\ket{\psi_0}$ could be the unentangled state $\ket{00\cdots 0}$. Then the complexity $\mathcal{C} (\ket \psi )$ is defined as the minimum number of simple gates (meaning they act on some $O(1)$ number of degrees of freedom) needed to take $\ket {\psi_0}$ to $\ket {\psi}$. The exact value of the complexity is, of course, dependent on details such as what gates are allowed, what the tolerance is, and so on. However, the qualitative behavior does not depend on these details: the complexity, in general, grows linearly with time for a large amount of time after the entanglement entropy saturates. 

\begin{figure}[tbp]

\centering
 \includegraphics[width=0.4\textwidth]{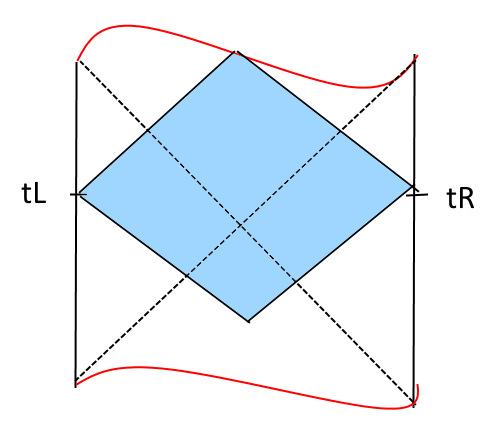}

 \caption{The Wheeler-de Witt patch of a two-sided AdS black hole. It is defined as the bulk domain of dependence for a slice between $t_L$ and $t_R$, shaded in light blue }\label{fig:WdWAdS}
\end{figure}

This line of reasoning led to the complexity-volume conjecture, which says that the CFT quantity dual to the maximum-volume slice is the complexity of the CFT state. Issues related to various ad-hoc factors in the complexity-volume conjecture led to the complexity action conjecture. Consider a AdS two-sided black hole. Then the complexity-action conjecture posits that the complexity of the state CFT state $\ket{\psi(t_L,t_R)}$ is given by \cite{Brown}
$$\mathcal{C} = \frac{\mathcal{I}_{WdW}}{\pi \hbar},$$
where $\mathcal{I}_{WdW}$ is the action on the Wheeler-de Witt patch, which is defined as the bulk domain of dependence of any spacelike surface between $t_L$ on the left boundary and $t_R$ on the right boundary. See Figure~\ref{fig:WdWAdS}. The authors of \cite{Brown} argued that is natural for the complexity to be normalized so that the constant of proportionality is $\frac{1}{\pi \hbar}$. In this way, when one fixes the normalization for one particular black hole, it is such that all other black holes saturate the upper limit on the rate of computation. 

However, cosmological observations indicate that our Universe is not AdS (which requires $\Lambda < 0$), but in fact has $\Lambda >0$. However, it is believed that holography holds in more general spacetimes than AdS space \cite{Bousso1,Bousso2}. A considerable amount of recent work has been done in trying to understand the structure of holography in general spacetimes. It is believed that the details of the holographic theory are encoded on a \textit{holographic screen}, which is a codimension-1 surface foliated by marginally trapped (or marginally anti-trapped) codimension-2 surfaces known as \textit{leaves}. (Recall that a codimension-2 surface will have two orthogonal null congruences, labelled by $k$ and $\ell$. The surface is called \textit{marginal} if one of the expansions, say $\theta_k$, vanishes. It is called \textit{marginally trapped} if $\theta_\ell <0$, and \textit{marginally anti-trapped} if $\theta_\ell >0$.) Recently, it was shown that holographic screens obey an area law \cite{BE1,BE2,BE3}.

The proposal of \cite{SW, HGST} is that, in general spacetimes, the holographic description of this gravity theory in the bulk lives on this holographic screen. In particular, \cite{SW} propose an analogy with the Ryu-Takayanagi formula for gravity in AdS spacetime. Specifically, if $\sigma(t)$ is a leaf of the holographic screen and $\Gamma \subset \sigma(t)$ is a subset of the leaf, then the entanglement entropy of the $\Gamma$ is given by 
$$S(\Gamma) = \frac{\text{Area}(E_\Gamma)}{4 G_N},$$
where $E_\Gamma$ is the extremal-area codimension-2 surface that is anchored on $\Gamma$: $\partial E_\Gamma = \partial \Gamma$.

\begin{figure}[tbp]

\centering
 \includegraphics[width=0.2\textwidth]{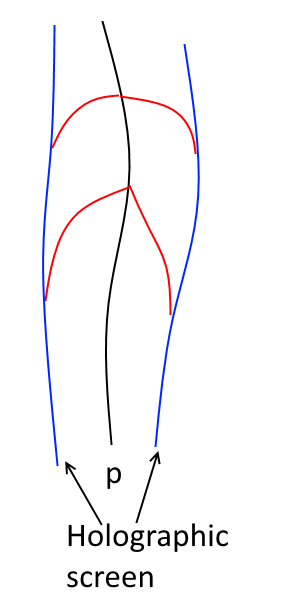}

 \caption{The construction of a holographic screen. At a point $\tau$ on the line $p(\tau)$, we follow the past null geodesics until they have zero expansion. Continuing in this way for the entire curve $p$, we have a codimension-1 surface that is foliated by leaves $\sigma(t)$.}\label{fig:HoloScreen}
\end{figure}

\begin{figure}[tbp]

\centering
 \includegraphics[width=0.4\textwidth]{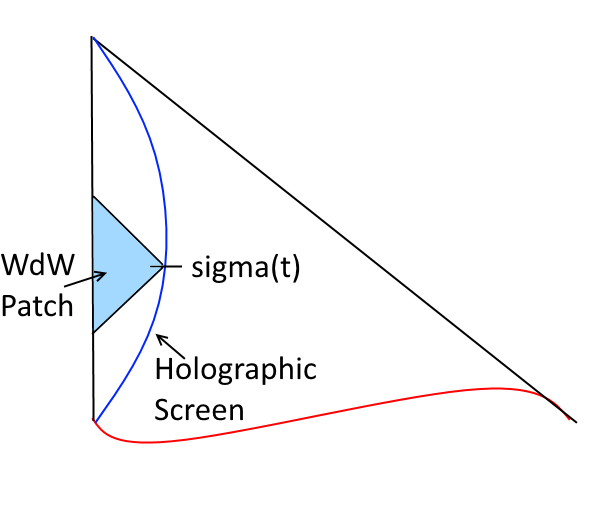}

 \caption{The Wheeler-de Witt patch for a general spacetime. It is the domain of dependence of a slice that is interior to a leaf $\sigma(t)$ of the holographic screen, shaded in light blue. }\label{fig:WdWFRW}
\end{figure}

The purpose of this paper is to examine the holographic complexity conjectures to more general spacetimes, in the same spirit as the hologaphic entropy conjecture for the holographic screens. Specifically, we study the same quantity (the action of the Wheeler-de Witt patch over $\pi \hbar$), but modify the WdW patch so that it is the domain of dependence of the interior of the leaf $\sigma(t)$. See Figure~\ref{fig:WdWFRW}.  We also examine the behavior of maximal-volume slices. We begin by reviewing the construction of the holographic screens.

\section{Holographic Screens}
We recall the construction of holographic screens. First, we pick some timelike path $p(\tau)$ through the spacetime. At each $\tau$, we fire a congruence of null geodesics in the past direction from $p(\tau)$. When the expansion parameter $\theta$ on this congruence reaches $0$, at some point in the past, this will be the location of the leaf of the holographic screen. By doing this for all values of $\tau$, we construct a codimension-one surface that is foliated by ``leaves" that have one expansion parameter vanishing. See Figure~\ref{fig:HoloScreen}. In the case of AdS space, as we show below, this reduces to the conformal boundary. 

\subsection{AdS Space}

First, we analyze AdS spacetimes. Consider, for example, AdS$_3$ in global coordinates, with the AdS scale set to 1. The metric is given by 

$$ds^2 = -(1+r^2)dt^2+\frac{dr^2}{1+r^2} + r^2 d \phi^2.$$
In these coordinates, $r$ ranges from $0$ to $\infty$, with the conformal boundary located at $r = \infty$. We choose our timelike path to simply be the line $r=0$. We now consider radial geodesics fired into the past from $r=0$. Because $\partial_t$ is a Killing vector, the quantity 
$$E = g_{tt} k^t = -(1+r^2) (k^t) $$
is conserved, and $E>0$ since the geodesics are past-directed. (We have defined $k^\alpha \equiv \frac{d x^\alpha}{d \lambda}.)$ Because the geodesic is null, we must have that 
$$-(1+r^2) (k^t)^2+ \frac{(k^r)^2}{1+r^2} =0$$
so that 
$$k^t = \frac{E}{1+r^2}, \text{	} k^r = E.$$
One can readily verify that $k^\alpha \nabla_\alpha k_\beta =0$ so this does indeed satisfy the geodesic equations. We compute the expansion parameter to find
$$\theta = g^{\alpha \beta} \nabla_{\alpha} k_\beta = \frac{E}{r}.$$
Therefore, we see that the expansion $\theta$ of this null congruence goes to 0 only at the boundary, $r \rightarrow \infty$. The holographic screen for AdS spacetime is the conformal boundary. Moreover, it is clear that the holographic screen for AdS will always be the conformal boundary, regardless of our choice for our timelike path $p(\tau)$ through the spacetime. 

\subsection{FRW Cosmologies} 

Next, we examine FRW cosmologies. The metric for this is given by 
$$ds^2 = -dt^2 + a^2(t) \left ( \frac{dr^2}{1-kr^2} + r^2 d \Omega_2 \right ),$$
where $a(t)$ is the scale factor. $k=0$ for a flat Universe, $+1$ for a closed Universe, and $-1$ for an open Universe. The equations governing the evolution of $a$ are the Friedmann equations
$$\frac{\dot a^2}{a^2}+ \frac{k}{a^2}= \frac{ 8 \pi \rho}{3},$$
$$ \dot \rho = -3 \frac{ \dot a} {a} (\rho + P).$$
We now find the holographic screen for the FRW metric. Again, we pick our path $p(\tau)$ through the spacetime to be the line $r=0$. We now study past directed null geodesics. The geodesic equation 
$$ \frac{d k^t}{d \lambda} + \Gamma^\alpha _{\beta \gamma} k^\beta k^\gamma =0,$$
(where $\lambda$ is an affine parameter) gives 
$$\frac{d k^t}{d \lambda} + \frac{a \dot a}{1-kr^2} (k^r)^2 =0.$$
Meanwhile, the geodesic is null so that $k^\alpha k_\alpha =0$ or 
$$(k^r)^2 = \frac{1-kr^2}{a^2} (k^t)^2$$
so that 
$$\frac{d k^t}{d \lambda} + \frac{ \dot a}{a} (k^t)^2 =0.$$
This equation is solved by $k^t = -\frac{\text{const}}{a}$. We calculate the expansion of this congruence of null geodesics. The result is 
$$\theta = g^{\alpha \beta} \nabla_{\alpha} k_{\beta}=\frac{2 \text{const}( \sqrt{1-kr^2}-r\dot a ) }{r a^2}.$$
The holographic screen will be at time $t$ and at the radius when $\theta=0$, which is given by
$$ r = \frac{1}{\sqrt{\dot a (t)^2 +k}}.$$

\section{Complexity-Action Conjecture in FRW Spacetimes}

We begin by computing the complexity given by the complexity-action conjecture \cite{Brown}. The contributions wil be the bulk Einstein-Hilbert term, a GHY term for spacelike boundaries, which we consider first. In section~\ref{sec:null_corner} below, we consider the null boundary terms and the corner terms. 

We will consider a flat ($k=0$) Universe with one component that has the equation of state
$$P= w \rho.$$
Then the scale factor behaves as (from the Friedmann equations) 
$$a(t) = c t^{\frac{2}{3(1+w)}}.$$
The $r$-coordinate of the holographic screen, at time $t_b$ is given by 
$$r = \frac{3(1+w)}{2} \frac{1}{c t_b ^{\frac{-1-3w}{3(1+w)}}}.$$
We define the scaled coordinates 
$$\rho = \frac{2}{3(1+w)}c t_b^{\frac{-1-3w}{3(1+w)}} r,$$
$$\eta =  \frac{2}{3(1+w)} \left ( \left ( \frac{t}{t_b}\right )^{\frac{1+3w}{3(1+w)}} -1 \right ).$$
The metric then becomes 
$$ds^2 = \frac{9 (1+w)^2 t_b^2}{4} \left (\frac{1+3w}{2} \eta + 1 \right )^{\frac{4}{1+3w}}[-d \eta^2 + d \rho^2 + \rho^2 d \Omega_2^2].$$
The leaf of the holographic screen is at $\eta =0,$ $\rho=1$. We first consider the upper half of the WdW patch, the part with $\eta \geq 0$. 

\begin{figure}[tbp]

\centering
 \includegraphics[width=\textwidth]{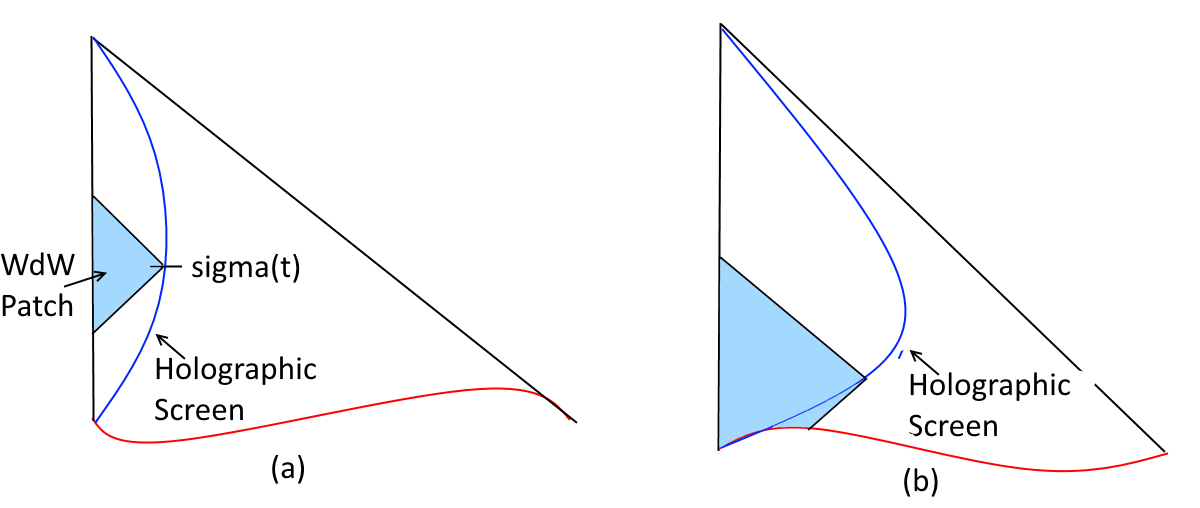}

 \caption{The Wheeler-de Witt patch for a FRW spacetimes. In some cases, (a) the WdW patch does not touch the Big Bang singularity. In others, (b), it touches the singularity and there will be a GHY boundary term associated to this boundary of the WdW patch in the singularity. }\label{fig:WdWFRWCases}
\end{figure}

Inward, future-directed geodesics from the leaf satisfy $\rho=1-\eta$. The inward, past-directed geodesics from the leaf have $\rho = 1+ \eta$. Under certain conditions, these geodesics will intersect the big bang singularity, $t=0$. We do this by linearly extrapolating back to find the value of $\rho$ at $t=0$. If $\rho>0$ at this point, it will intersect the singularity. If $\rho<0$, the WdW patch terminates at some strictly positive value of $t$. See Figure~\ref{fig:WdWFRWCases}. When $t=0$, 
$$\eta = - \frac{2}{3(1+w)}$$
so that 
$$\rho = 1+ \eta = 1 - \frac{2}{3(1+w)}=\frac{1+3w}{3(1+w)}.$$
Thus, if $w < -1/3$, the WdW patch does not intersect the singularity, while if $w>-1/3$ it does. 

 We first consider the upper half of the WdW patch, the part with $\eta \geq 0$. The bulk contribution to the gravitational action from this region is given by 
 $$\mathcal{I}_{up}=\frac{1}{16 \pi G_N} \int_0^1 d \eta \int_0^{1-\eta} d \rho \sqrt{-g} R \int d\Omega_2,$$
where $R$ is the Ricci scalar. Using Mathematica, we find that 
$$ R= 9\frac{ 2^{\frac{8}{3 w+1}-1} (\eta +3 \eta  w+2)^{-\frac{8}{3
   w+1}} \left(1-16^{\frac{1}{-3 w-1}} (\eta +3 \eta 
   w+2)^{\frac{4}{3 w+1}-2} \left((\eta +3 \eta  w+2)^2+4 \rho ^2 (6
   w-5)\right)\right)}{\rho ^2 t^2 (w+1)^2}.$$
We then perform the integration to obtain
\begin{multline}\mathcal{I}_{up} = \frac{t^2}{4 G_N} 2^{-\frac{6}{3 w+1}-2} 3^{\frac{2}{3 w+1}-1}
   (w+1)^{\frac{2}{3 w+1}} [(3 w-5) (3 w+2) (3 w+4) (3 w+7)]^{-1} \\ \cdot (2^{\frac{6}{3 w+1}+3}  
   \left(-162
   w^4-675 w^3-531 w^2+513 w+515\right) (3 w+3)^{\frac{6 w}{3
   w+1}}+ \\ 
   81 (w+1)^3 \left(5 (w+1) (3 w-1) (3 w+2) (3
   w+3)^{\frac{4}{3 w+1}}+16^{\frac{1}{3 w+1}} (3 w-5) (3 w+4) (3
   w+7)\right)).
\end{multline}
We now consider the lower half of the WdW patch. We begin by considering the case where it does not intersect the Big Bang singularity. We fire past-directed light rays from the leaf, which satisfy $\rho = 1 + \eta$, so they will hit $\rho=0$ when $\eta = -1$. The bulk contribution to the gravitational action for the lower WdW patch is then given by 
 $$\mathcal{I}_{low}=\frac{1}{16 \pi G_N} \int^0_{-1} d \eta \int_0^{1+\eta} d \rho \sqrt{-g} R \int d\Omega_2.$$
We perform this integral to obtain 
\begin{multline}\mathcal{I}_{low} = \frac{t_b^2}{4G_N} [(3 w-5) (3 w-1) (3 w+2) (3 w+4) (3
   w+7)]^{-1} 3\cdot2^{-\frac{6}{3 w+1}-2} (w+1) (8^{\frac{2}{3 w+1}+1}
   (3 w-1) \\ \cdot \left(3 w \left(36 w^3+84
   w^2+ w-86\right)-85\right)  -16^{\frac{1}{3 w+1}} (3 w-5) (3 w+4) (3
   w+7) (1-3 w)^{\frac{2}{3 w+1}+3}+ \\ 5 (w+1) (3 w+2) (1-3
   w)^{\frac{6}{3 w+1}+4}).
\end{multline}
The total bulk gravity action for the WdW patch is, then, (for $w<-1/3$):
\begin{multline}\mathcal{I}^{WdW}_{bulk,gravity} = \mathcal{I}_{low}+\mathcal{I}_{up}=   \\ \frac{t_b^2}{4G_N} [2^{-\frac{6}{3 w+1}-2} (\frac{1}{3w+1}(9 (8^{\frac{2}{3
   w+1}+1} (3 w-1) \left(3 w \left(36 w^3+84
   w^2+w-86\right)-85\right) \\ -  16^{\frac{1}{3 w+1}} (3 w-5) (3 w+4) (3
   w+7) (1-3 w)^{\frac{2}{3 w+1}+3}+5 (w+1) (3 w+2) (1-3
   w)^{\frac{6}{3 w+1}+4}) (w+1))\\
   +9^{\frac{1}{3 w+1}}
   (2^{\frac{6}{3 w+1}+3} \left(-162 w^4-675 w^3-531 w^2+513
   w+515\right) (3 w+3)^{\frac{6 w}{3 w+1}}+ \\
   81 (w+1)^3 \left(5 (w+1)
   (3 w-1) (3 w+2) (3 w+3)^{\frac{4}{3 w+1}}+16^{\frac{1}{3 w+1}} (3
   w-5) (3 w+4) (3 w+7)\right)) (w+1)^{\frac{2}{3
   w+1}})] \\
   \cdot [3 (3 w-5) (3 w+2) (3 w+4) (3 w+7)]^{-1}.
\end{multline}

Next, we consider the case where the WdW patch does intersect the Big Bang singularity, which is the case when $w > -1/3$. As can be seen from Figure~\ref{fig:WdWFRWCases}, there will be a Gibbons-Hawking-York boundary term associated to the boundary, where the WdW patch intersects the singularity. The upper bulk WdW patch remains unchanged, while the action associated to the lower WdW patch has to be modified -- we need to integrate $\eta$ from its value at the initial singularity (when $t=0$) to $0$. Thus, we have 
 $$\mathcal{I}_{low}=\frac{1}{16 \pi G_N} \int^0_{- \frac{2}{3(1+w)}} d \eta \int_0^{1+\eta} d \rho \sqrt{-g} R \int d\Omega_2.$$
We preform this integral on Mathematica to obtain the total bulk gravity action 
\begin{multline}\mathcal{I}^{WdW}_{bulk,gravity} = \mathcal{I}_{low}+\mathcal{I}_{up}=\frac{t_b^2}{4G_N} 
\frac{3}{2} \{  (\frac{4 (w+1) \left(3 w \left(36 w^3+84
   w^2+w-86\right)-85\right)}{(3 w-5) (3 w+2) (3 w+4) (3
   w+7)}- \\ \frac{4^{\frac{1}{3 w+1}+1} 9^{\frac{1}{-3 w-1}} w
   \left(\frac{1}{w+1}\right)^{\frac{2}{3 w+1}}}{3
   w+2}- \\ \frac{9^{-\frac{3}{3 w+1}-1} 64^{\frac{1}{3 w+1}} (3 w-1) (3
   w (3 w (3 w (9 w (2 w+7)+37)-101)-179)-100)
   \left(\frac{1}{w+1}\right)^{\frac{6}{3 w+1}}}{(3 w-5) (3 w+4) (3
   w+7)} )+ \\ [2^{-\frac{6}{3 w+1}-2} 3^{\frac{2}{3 w+1}-1}
  (2^{\frac{6}{3 w+1}+3} \left(-162 w^4-675 w^3-531 w^2+513
   w+515\right) (3 w+3)^{\frac{6 w}{3 w+1}}+\\ 81 (w+1)^3 \left(5 (w+1)
   (3 w-1) (3 w+2) (3 w+3)^{\frac{4}{3 w+1}}+16^{\frac{1}{3 w+1}} (3
   w-5) (3 w+4) (3 w+7)\right)) (w+1)^{\frac{2}{3 w+1}}] \\ [(3
   w-5) (3 w+2) (3 w+4) (3 w+7)]^{-1} \}.
   \end{multline}
Now, the boundary term is given by 
$$\mathcal{I}_{GHY} = \frac{1}{ 8 \pi G_N} \int d^3 x \sqrt{h} K,$$
where $h$ is the induced metric on the boundary, while $K$ is the trace of its extrinsic curvature. We consider a $\eta = \text{const}$ slice.  The normal vector is given by $n_\eta = \sqrt{g_{\eta \eta}}$, all other components 0. We can compute $K_{\mu \nu} = \nabla_\mu n_\nu$ and then take the trace $K = K^\mu_{\text{		}\mu}$. The $\rho$ integral ranges from $0$ to $1+\eta$. Evaluating the geometric quantities on Mathematica, we obtain 
\begin{multline}\mathcal{I}_{GHY} = \frac{1}{ 2 G_N} \int_0^{1+\eta} d \rho \rho^2 \left[-9 t_b^2  (w+1)^2 (\eta  \frac{1+3w}{2}+1)^{\frac{4}{3
   w+1}-1}\right ] \\
   =\frac{-9t_b^2}{2G_N}(w+1)^2 (\eta  \frac{1+3w}{2}+1)^{\frac{4}{3
   w+1}-1} \int_0^{1+\eta} d \rho \rho^2 \\
   = \frac{-3 t_b^2}{2G_N}(w+1)^2 (\eta  \frac{1+3w}{2}+1)^{\frac{4}{3
   w+1}-1} (1+\eta)^3.
\end{multline}
This is for a surface at $\eta$ -- the term of interest is obtained as we go to the big bang, i.e., by taking the limit $\eta \rightarrow \eta_{\text{Big Bang}} = -\frac{2}{3(1+w)}$. Taking this limit, we obtain
$$\mathcal{I}_{GHY} = \frac{-3 t_b^2}{2G_N}(w+1)^2  (\frac{2-3w}{3+3w})^{\frac{4}{3
   w+1}-1} (\frac{1+3w}{3+3w})^3.$$

Notice how both the GHY term and the bulk terms are both proportional to $t_b^2$. This is in contrast to the case of black holes in AdS, where generically, the WdW action is linear in $t$. A possible (very schematic) explanation for this is follows. Roughly speaking, in any system we expect that the rate of change of the complexity to be proportional to its entropy \cite{LPiTP}:
$$\frac{d \mathcal{C}}{dt} \sim S. $$
In the case of AdS/CFT, the entropy is a constant -- as we evolve forward in time in the boundary Hilbert space, we are in a constant time-slice of the CFT. The entropy is roughly the area of a time slice of the boundary (by the holographic principle), and this is of course unchanging. Therefore, we would expect that 
$$\mathcal{C}_{AdS} \propto t.$$
However, in the case of FRW spacetimes, as we evolve forward in time, the Hilbert space changes. The dimension of the Hilbert space is roughly given by the area of the leaf. Hence, the complexity at boundary time $t$ should grow as 
$$\frac{d \mathcal{C}}{dt} \sim S \sim \frac{\text{Area}[\sigma(t)]}{G_N},$$
where as usual $\sigma(t)$ is a leaf of the holographic screen at time $t$. Now, one also has
$$\text{Area}[\sigma(t)] \sim \left( \frac{a(t)}{\dot a(t)} \right)^2 \sim t^2.$$
Thus, we would expect 
$$\frac{d \mathcal{C}}{dt} \sim t^{O(1)},$$
which is what we found from the bulk computation. The relationship between the Hilbert spaces of each of the leaves and gravity theory is not yet established. It is possible that these effects will yield an $O(1)$ exponent for $d \mathcal{C}/dt.$

\subsection{Null Boundary and Corner Terms}
\label{sec:null_corner} 

In addition to the bulk Einstein-Hilbert term, and the spacelike boundary GHY term, we also must consider null boundary terms and corner terms. This was first worked out in \cite{Corner}, and we follow their procedure, which we summarize in Appendix \ref{app:A}. 

For a given spacetime region $W$ with a null component $N$ of the boundary $\partial W$, its boundary and corner terms are given by

\begin{multline}
 \mathcal{I}_N = \frac{sgn(N)}{8 \pi G_N} \int_N d \lambda d \theta \sqrt{\gamma} \kappa + \frac{sgn(N)}{8 \pi G_N} \int_N d \lambda d \theta \sqrt{\gamma} \Theta \log ( l_c \abs{\Theta}) \\ + \frac{1}{8 \pi G_N} \int_{B_1} d \theta \sqrt{\gamma} a_1 +\frac{1}{8 \pi G_N} \int_{B_2} d \theta \sqrt{\gamma} a_2.
 \end{multline}

In the above equation, $sgn(N)$ is +1 if $N$ is to the future of $W$, and -1 if it is to the past of $W$. Also, $\lambda$ represents a parameterization of the null generators, and the coordinates $\theta$ label the different null generators of $N$. Meanwhile, $B_1$ and $B_2$ are the endpoints of the component $N$. The null geodesics will satisfy the equation 
$$ k^\mu \nabla_\mu k^\nu = \kappa k^\nu.$$
If $\lambda$ is an affine parameterization, then $\kappa$ is of course 0. $\Theta$ is the expansion of $N$, while $\gamma$ is the transverse metric. Finally, $l_c$ is an undetermined parameter in the counterterm. As we summarize in Appendix \ref{app:A}, this is independent of the parameterization of the null geodesics.

For our case, we first consider the geodesics to have an affine parameterization, which eliminates the first term. Specifically, we choose our geodesics to be parameterized such that the tangent vectors are 
$$ k =  \left (\frac{1}{a}, -\frac{1}{a^2}, 0, 0 \right )$$
for the upper sheet and 
$$ m= \left (\frac{1}{a}, \frac{1}{a^2}, 0, 0 \right)$$
for the lower sheet. 

For the second term, we first consider the upper part of the boundary. For an FRW universe with equation of state $P= w \rho$, the scale factor evolves as 
$$a = t^b, \text{	} b \equiv \frac{2}{3(1+w)}.$$
 
 The holographic screen is at location $t_b$, $r = \frac{1}{b t^{b-1}}$. The geodesic that composes the upper portion of the boundary will satisfy
 $$r(t) = \frac{1}{b t_b^{b-1}}+ \frac{-t^{1-b}+t_b^{1-b}}{1-b}.$$
 Meanwhile, the expansion is 
 $$\Theta = \frac{-2 b t^{-b-1}}{r},$$
 while the transverse metric is 
 $$\sqrt{\gamma} = r^2 a(t)^2 d \Omega_2.$$
 We can rewrite the integral over $\lambda$ as an integral over $t$, integrating from $t_b$ to the point when $r=0$ along the null geodesic. This occurs when 
 
$$t = t_{max} = \left ( t_b^{1-b} +\frac{1-b}{b t_b^{b-1}} \right )^{\frac{1}{1-b}}.$$
The integral gives
   \begin{multline}\mathcal{I}_{surf}^{up} = \frac{1}{ 8 \pi G_N (b-1) b^2}  [2 t_b^{-b} (b \left(t_b t^b-b^2 t t_b^b\right) \log \left(\frac{2 (b-1) b^2 l_c t_b^b}{t
   \left(t_b t^b-b t t_b^b\right)}\right)-b^2 \left(b^2+1\right) t t_b^b+ \\  (b-1)^2 b^2 t t_b^b \,
   _2F_1\left(1,\frac{1}{b-1};\frac{b}{b-1};\frac{t_b^{1-b} t^{b-1}}{b}\right)+(b+1) t_b
   t^b)  ]^{t=t_{max}}_{t=t_b}. 
   \end{multline}

We now calculate the surface counterterm for the lower part of the boundary. In this case, the null geodesic is described by
$$r(t) = \frac{1}{b t_b^{b-1}}+ \frac{t^{1-b}-t_b^{1-b}}{1-b}.$$
The $t$ integral runs from $t_{min}$, the value of $t$ when the expression above for $r(t)$ reaches 0, to $t_b$. We have that 
$$t_{min} = \left (  t_b^{1-b} - \frac{1-b}{b t_b^{b-1}} \right)^{\frac{1}{1-b}}.$$
The resulting integral is

\begin{multline} \mathcal{I}_{surf}^{low} = \frac{-1}{8 \pi G_N} \frac{1}{(b-1) b^2} 2 t_b^{-b} [b^4 t t_b^b+b \left(b^2 t t_b^b-2 b t_b t^b+t_b t^b\right) \log
   \left(\frac{2 (b-1) b^2 l_c t_b^b}{t \left(b t t_b^b-2 b t_b t^b+t_b t^b\right)}\right)+ \\ b^2 t
   t_b^b-(b-1)^2 b^2 t t_b^b \, _2F_1\left(1,\frac{1}{b-1};\frac{b}{b-1};\frac{(2 b-1)
   t_b^{1-b} t^{b-1}}{b}\right)-2 b^2 t_b t^b-b t_b t^b+t_b t^b]_{t=t_{min}}^{t=t_b}.  \end{multline}

Meanwhile, we consider the corner terms. They are given by 
$$\mathcal{I}_{corner} = \frac{1}{8 \pi G_N} \int a d S,$$
where 
$$a = \pm \log \abs {\frac{ k \cdot m}{2}},$$
with $k$ and $m$ being the two normal vectors to the null sheets that join at the corner. The sign is determined is follows. If $W \subset J^+(N)$, and the corner is at the past end of $N$, or if $W \subset J^-(N)$, and the corner is at the future end of $N$, then the sign is positive. In all other cases, it is negative. In our case, the sign is negative. 
We know that
$$ k =  \left (\frac{1}{a}, -\frac{1}{a^2}, 0, 0 \right )$$
for the upper sheet and 
$$ m= \left (\frac{1}{a}, \frac{1}{a^2}, 0, 0 \right)$$
for the lower sheet. Thus, the corner term for the joint between the null sheet fired in towards the future, and the null sheet fired inwards towards the past is:

$$\mathcal{I}_{corner} = -\frac{1}{8 \pi G_N} \frac{ t_b^{2b}}{b^2 t_b^{2b-2}} \log \left [ \frac{1}{2} \abs {- \frac{1}{a^2} -\frac{1}{a^2} } \right ] 4 \pi = - \frac{1}{2 G_N} \frac{t_b^2}{b^2} \log \left [ \frac{1}{t_b^{2b}}   \right ] = \frac{1}{G_N} \frac{t_b^2}{b} \log(t_b).$$

\section{Holographic Screen Complexity in an FRW Spacetime undergoing a Transition}

We wish to calculate the holographic screen complexity for a flat FRW Universe undergoing a transition from a state of matter with one equation of state to another. For concreteness, we consider a Universe where the matter is a scalar field $\phi$, following section 3.2 of~\cite{HGST}. We consider a scalar field with potential 
$$V(\phi) = 1 - e^{-k(\phi-\phi_0)^2} + s (\phi -\phi_0) \tanh (p (\phi-\phi_0)).$$
%
We consider two values of the parameters: a ``steep" potential, with values
$$k = 5000, s = 0.01, p = 20, \phi_0 = 0.045,$$
and a ``broad" potential, with values
$$k = 25, s = 0.01, p = 2, \phi_0 = 0.5,$$
as in \cite{HGST}.
 The energy density of the scalar field is given by 
$$\rho = \frac{1}{2} \dot \phi^2 + V( \phi),$$
so that the Friedmann equation is 
$$\frac{1}{a} \frac{da}{dt} = \sqrt{\frac{8 \pi}{3}} \sqrt{\frac{1}{2} \dot \phi^2 + V( \phi)},$$
where an overdot denotes a derivative with respect to $t$. The equation of motion for $\phi$, meanwhile, is 
$$\frac{d^2 \phi}{d t^2}+3 \frac{\dot a}{a} \frac{d \phi}{dt}+ V'(\phi)=0,$$
where a prime denotes a derivative with respect to the scalar field $\phi$. Therefore, we have 
$$\frac{d^2 \phi}{d t^2}+3 \sqrt{\frac{8 \pi}{3}} \sqrt{\frac{1}{2} \dot \phi^2 + V( \phi)} \frac{d \phi}{dt}+ V'(\phi)=0.$$

\begin{figure}[tbp]
\centering
\begin{subfigure}{0.45\textwidth}
 \includegraphics[width=\textwidth]{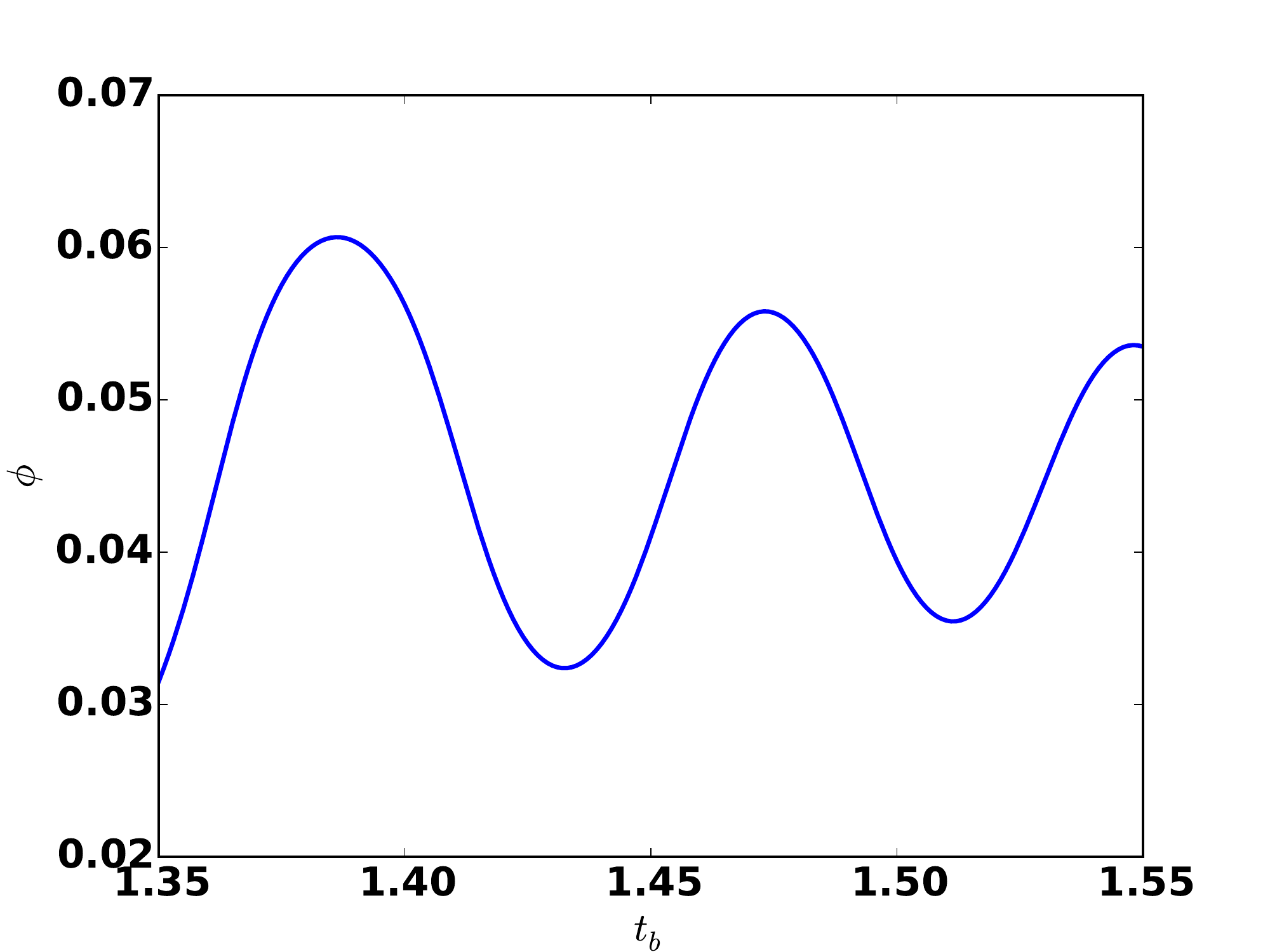}
 \caption{$\phi$ vs $t$}
\end{subfigure}
\begin{subfigure}{0.45\textwidth}
 \includegraphics[width=\textwidth]{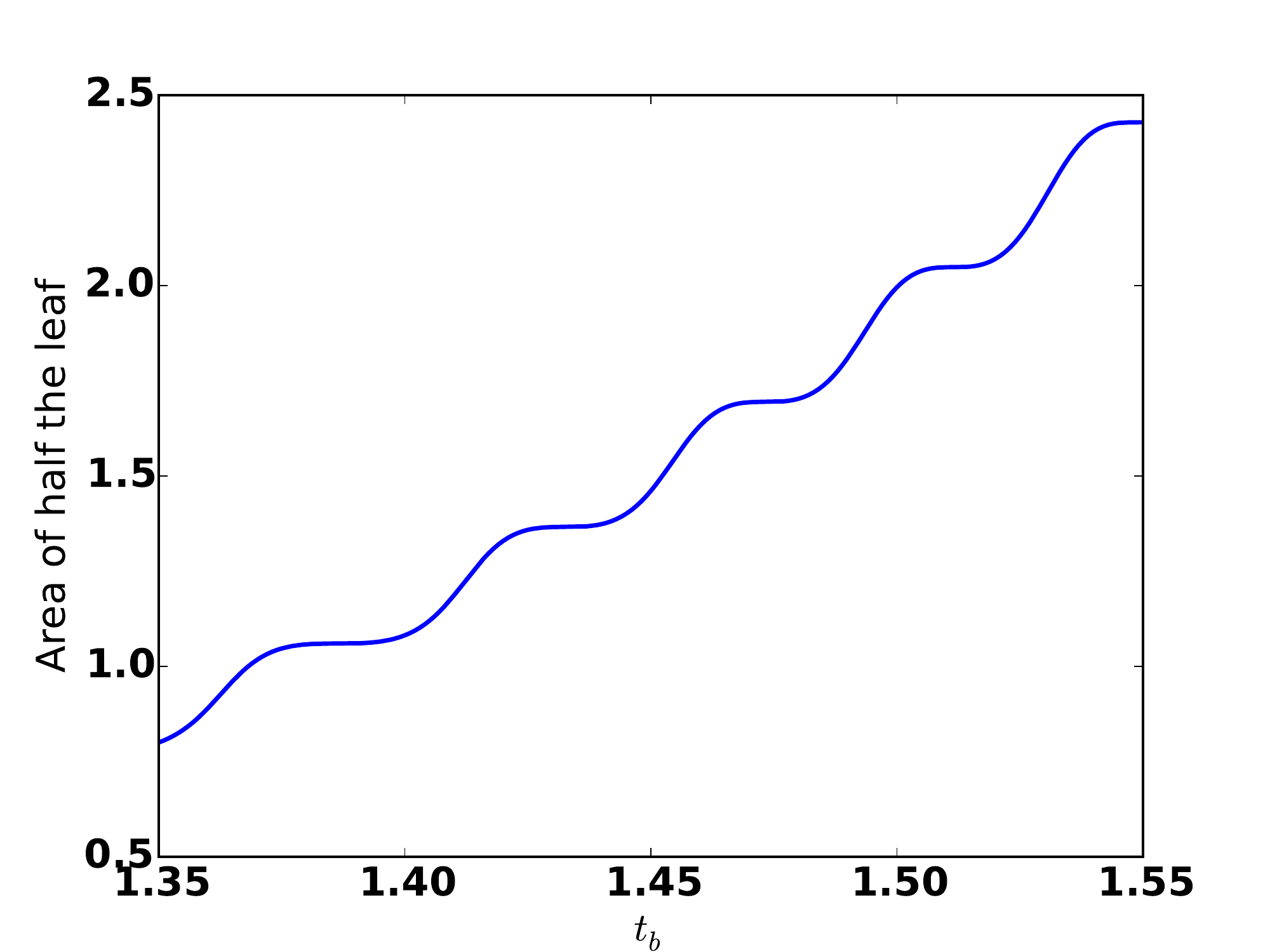}
 \caption{Area of the half leaf vs $t$}
\end{subfigure}

 \caption{Numerical solution of $\phi$, as well as the area of the half leaf, for the steep potential.}\label{fig:SteepSln}
\end{figure}

\begin{figure}[tbp]
\centering
\begin{subfigure}{0.45\textwidth}
 \includegraphics[width=\textwidth]{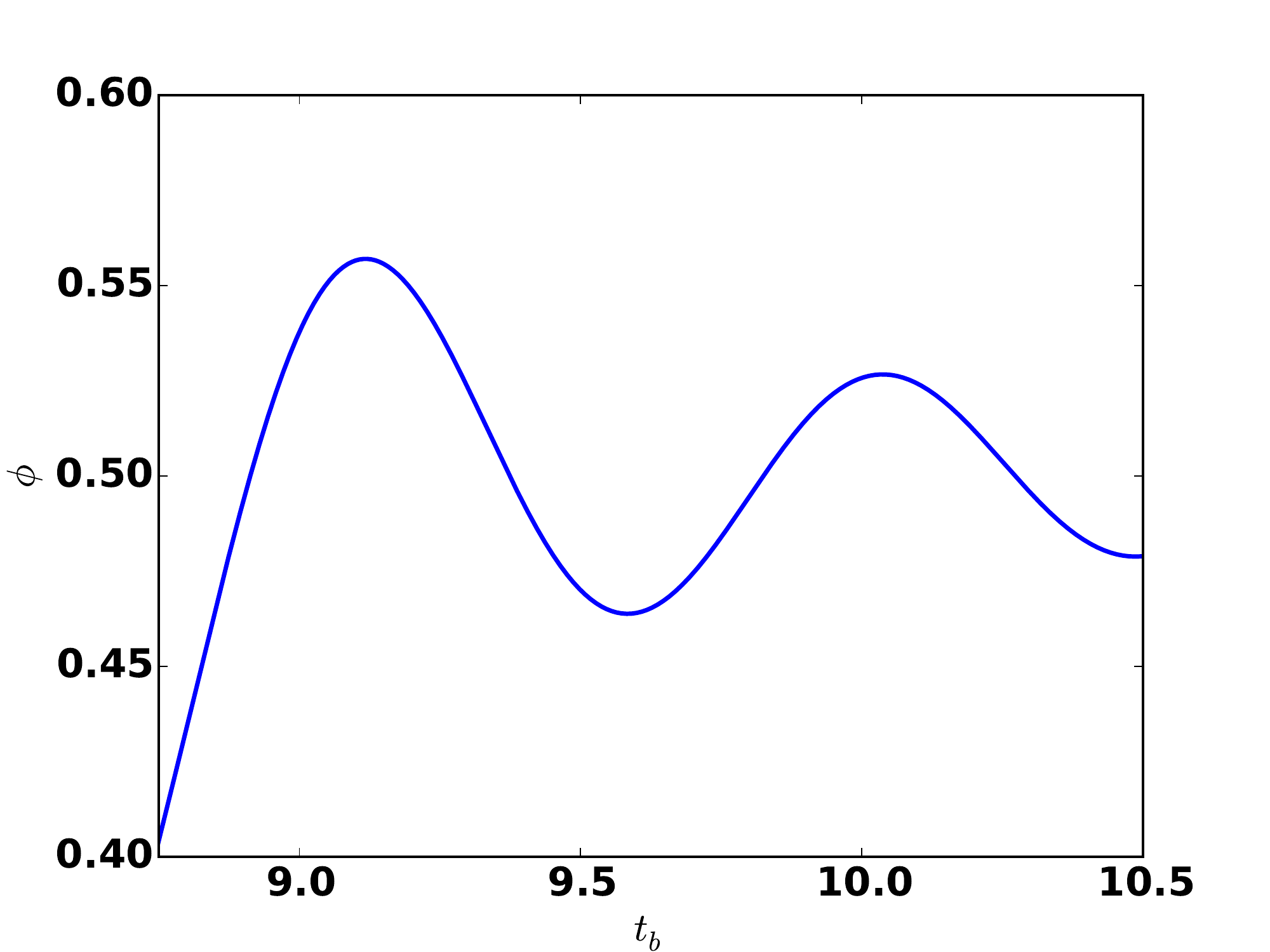}
 \caption{$\phi$ vs $t$}
\end{subfigure}
\begin{subfigure}{0.45\textwidth}
 \includegraphics[width=\textwidth]{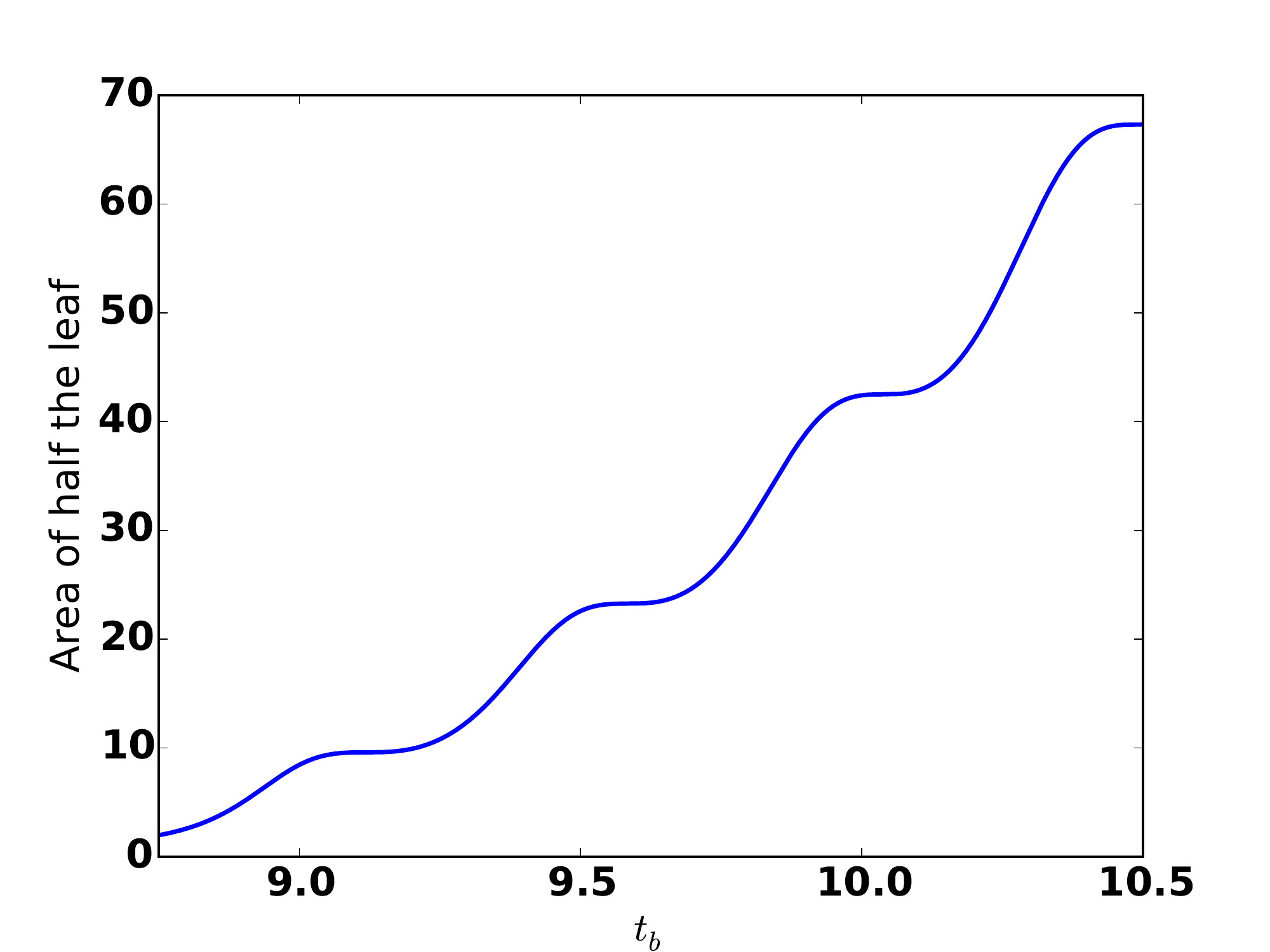}
 \caption{Area of the half leaf vs $t$}
\end{subfigure}

 \caption{Numerical solution of $\phi$, as well as the area of the half leaf, for the broad potential.}\label{fig:BroadSln}
\end{figure}

We integrate this numerically, together with the Friedmann equation. We plot $\phi$ as a function of $t$, as well as the area of half the holographic leaf, in Figure~\ref{fig:SteepSln} for the steep potential and \ref{fig:BroadSln} for the broad potential. In addition, we also plot the proper area of the leaf of holographic $\sigma(t)$. 

The Wheeler-de Witt patch associated to $\sigma(t_b)$ is given by 
$$0 \leq r \leq \tilde{r}(t), t_{low} \leq 0 \leq t_{up},$$
where $\tilde{r}(t)$ is given by 
$$r_{AH}(t) - \int_{t_b}^{t} \frac{dt'}{a(t')}$$ 
if $t<t_b$, and 
$$r_{AH}(t) + \int_{t_b}^{t} \frac{dt'}{a(t')}$$ 
if $t>t_b$. $t_{low}$ and $t_{up}$ are, of course, the points where $\tilde{r}(t)=0$. The Ricci scalar of the flat FRW metric
$$ds^2 = -dt^2 + a^2(t) (dr^2 +r^2 d \Omega_2^2)$$
is given by 
$$R = 6 \left [ \frac{\ddot a }{a} + \left( \frac{\dot a}{a} \right )^2 \right ].$$
The gravity action of the WdW patch is given by 
\begin{multline}\mathcal{I}_{grav} = \frac{1}{16 \pi} \int_{WdW} d^4 x \sqrt{-g} R =\frac{3}{8 \pi} \int d \Omega_2 \int_{t_{low}}^{t_{up}} dt \int_0^{\tilde{r}(t)}dr r^2 (\ddot{a} a^2 + \dot a ^2 a) \\ 
=\frac{1}{2}  \int_{t_{low}}^{t_{up}} dt  \tilde{r}(t)^3 (\ddot{a} a^2 + \dot a ^2 a).
\end{multline}
Meanwhile, the action of the scalar field is given by
$$\mathcal{I}_{sc} = \int_{WdW} d^4 x \sqrt{-g} \left ( \frac{1}{2} \dot \phi^2 - V(\phi) \right) = \frac{1}{3} \int d \Omega_2 \int_{t_{low}}^{t_{up}} dt \tilde{r}(t)^3 \left (\frac{1}{2}\dot \phi^2 - V(\phi) \right )$$
so that the total action on the WdW patch is given by
$$\mathcal{I}_{WdW} = \mathcal{I}_{grav} + \mathcal{I}_{sc}.$$


\begin{figure}[tbp]
\centering
\begin{subfigure}{0.45\textwidth}
 \includegraphics[width=\textwidth]{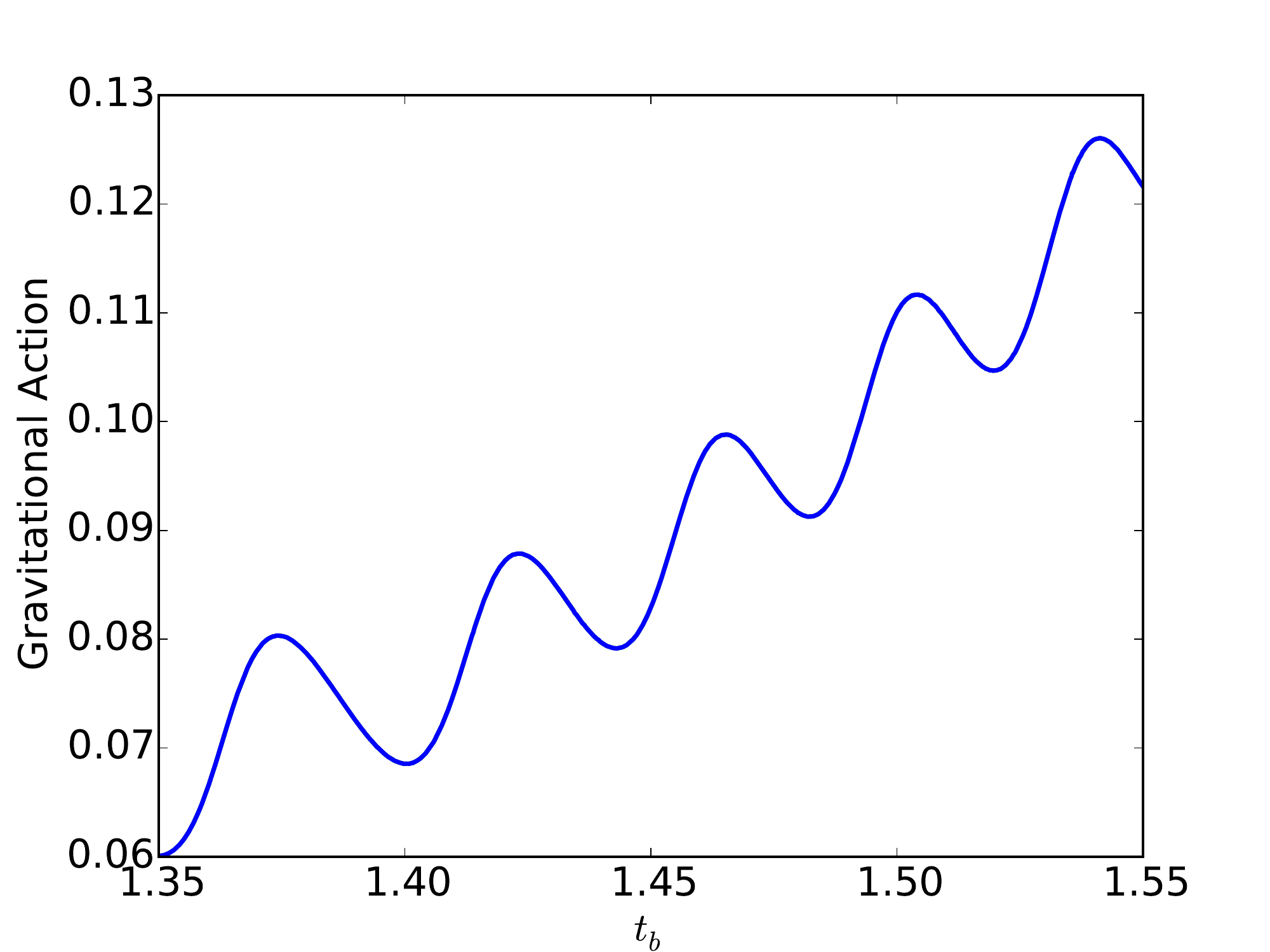}
 \caption{Gravitational WdW action for the steep potential}
\end{subfigure}
\begin{subfigure}{0.45\textwidth}
 \includegraphics[width=\textwidth]{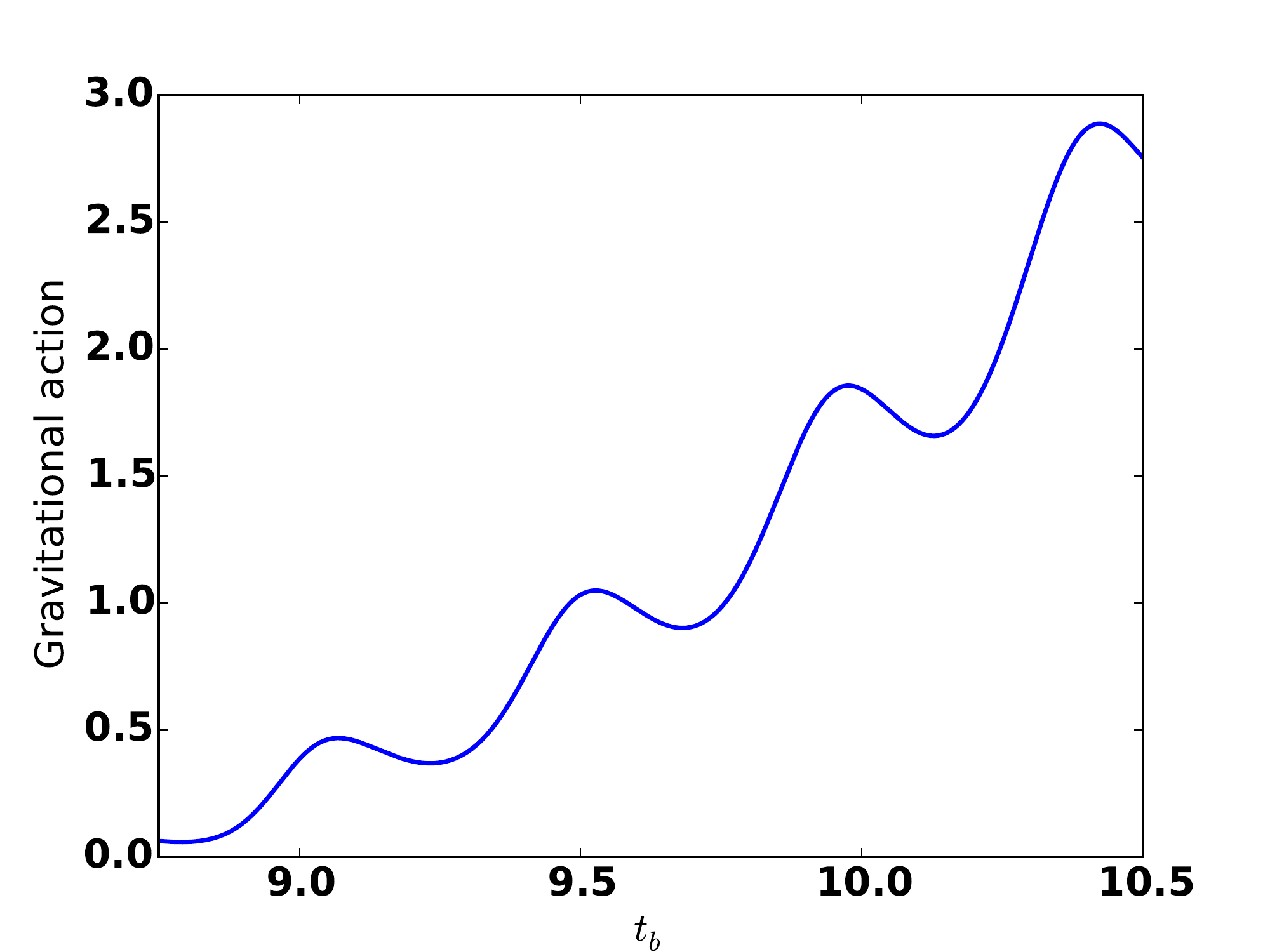}
 \caption{Gravitational WdW action for the broad potential}
\end{subfigure}
\begin{subfigure}{0.45\textwidth}
 \includegraphics[width=\textwidth]{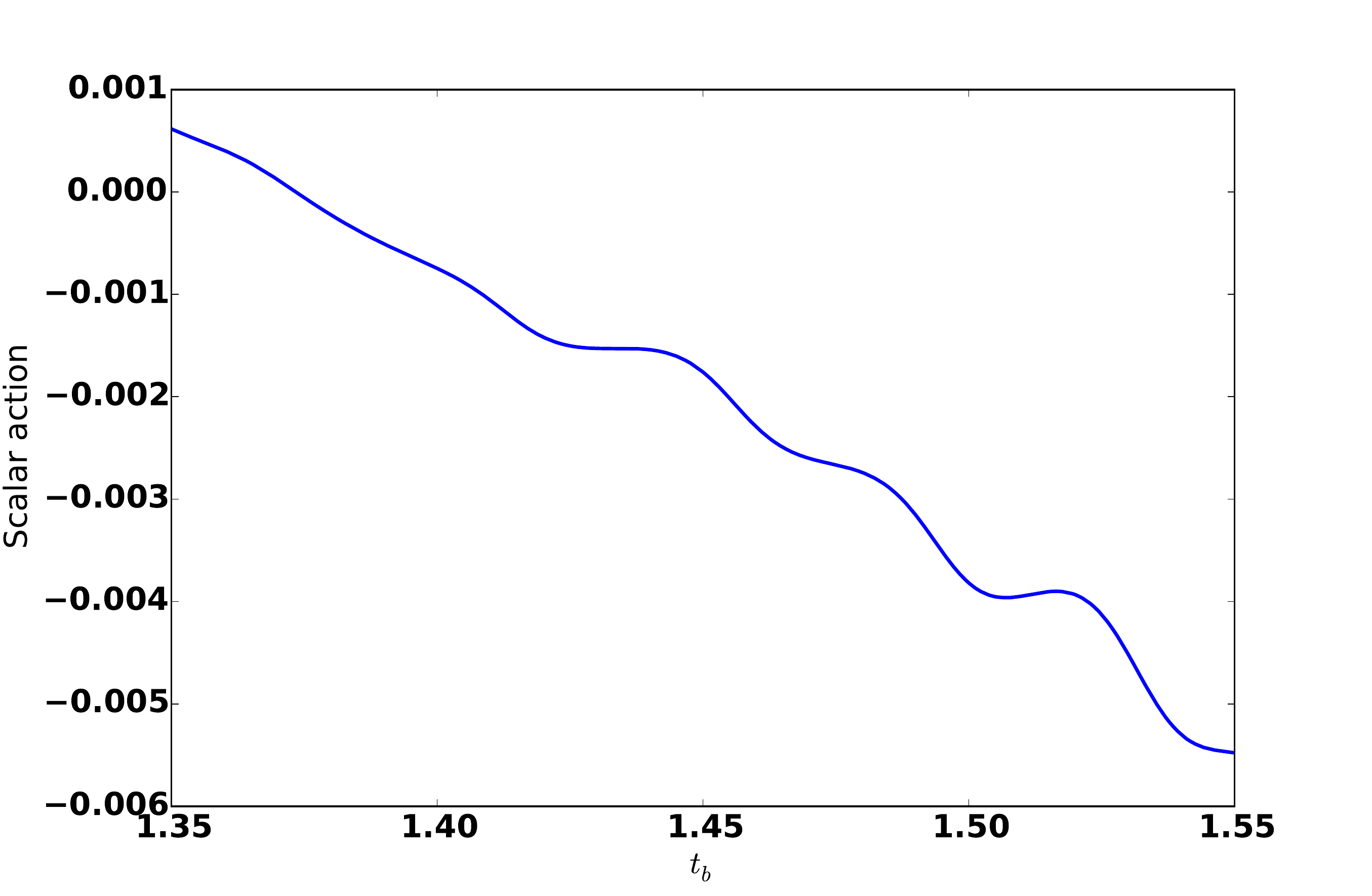}
 \caption{Scalar WdW action for the steep potential}
\end{subfigure}
\begin{subfigure}{0.45\textwidth}
 \includegraphics[width=\textwidth]{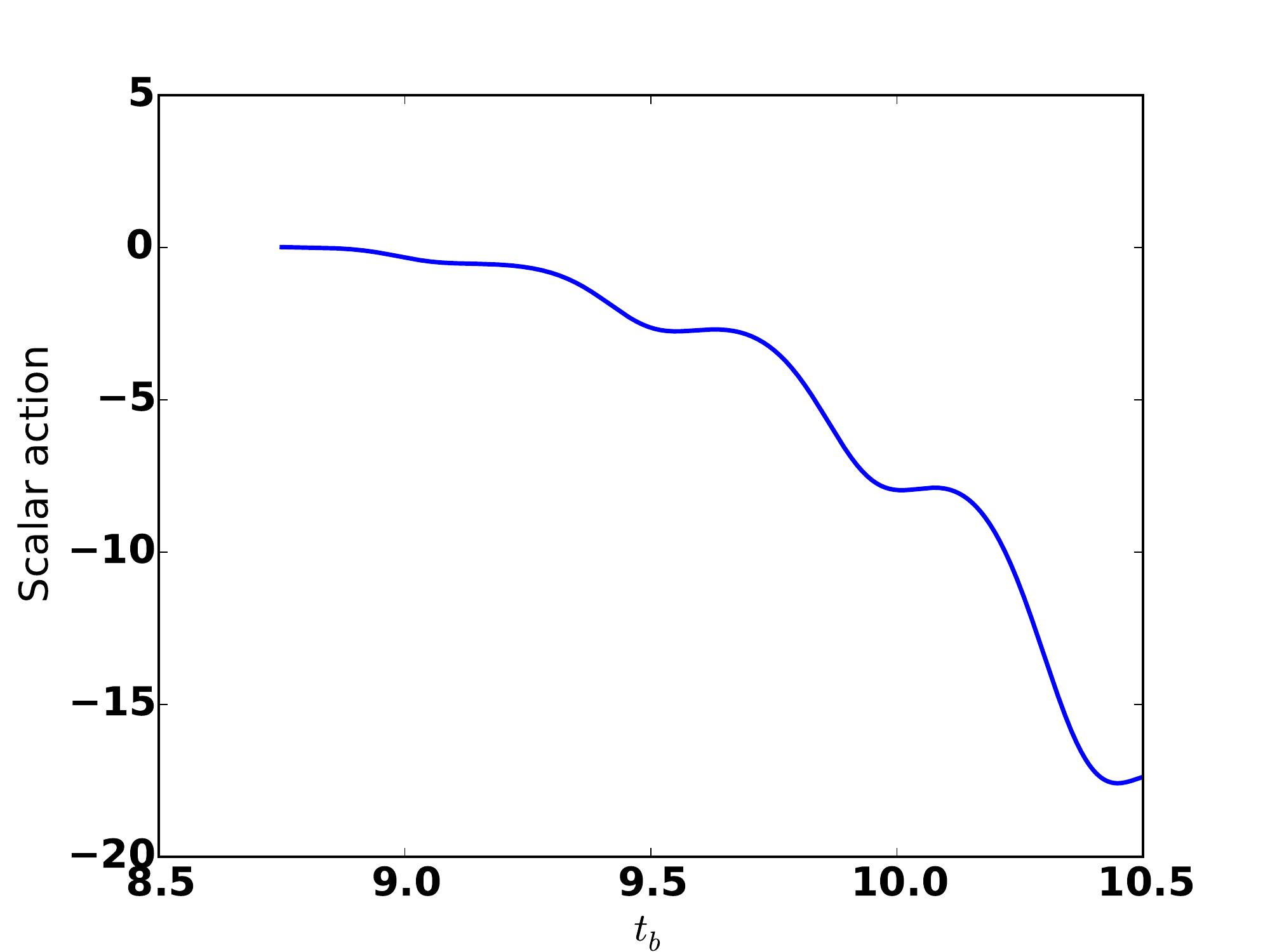}
 \caption{Scalar WdW action for the broad potential}
\end{subfigure}
 \begin{subfigure}{0.45\textwidth}
 \includegraphics[width=\textwidth]{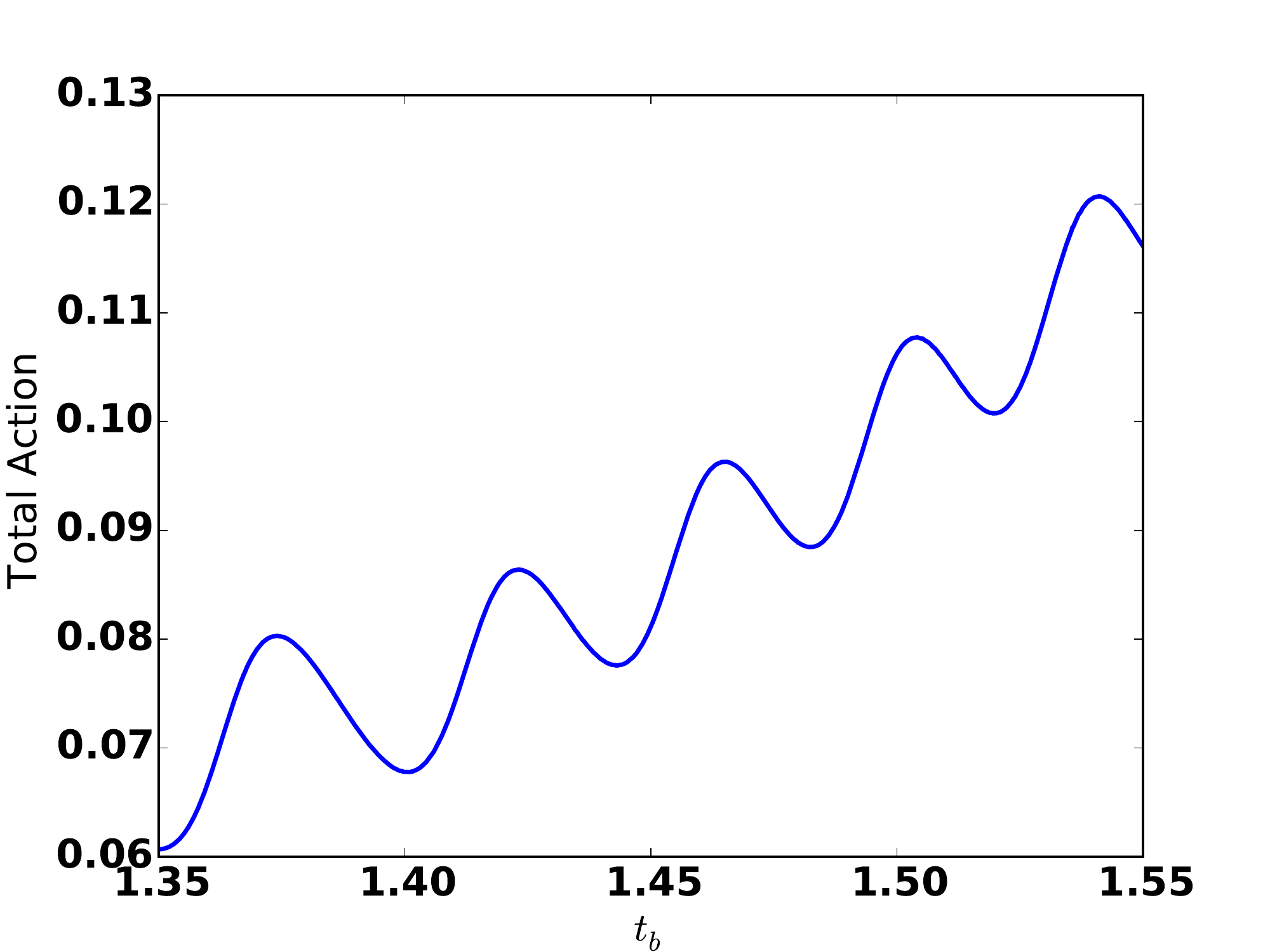}
 \caption{Total WdW action for the steep potential}
\end{subfigure}
\begin{subfigure}{0.45\textwidth}
 \includegraphics[width=\textwidth]{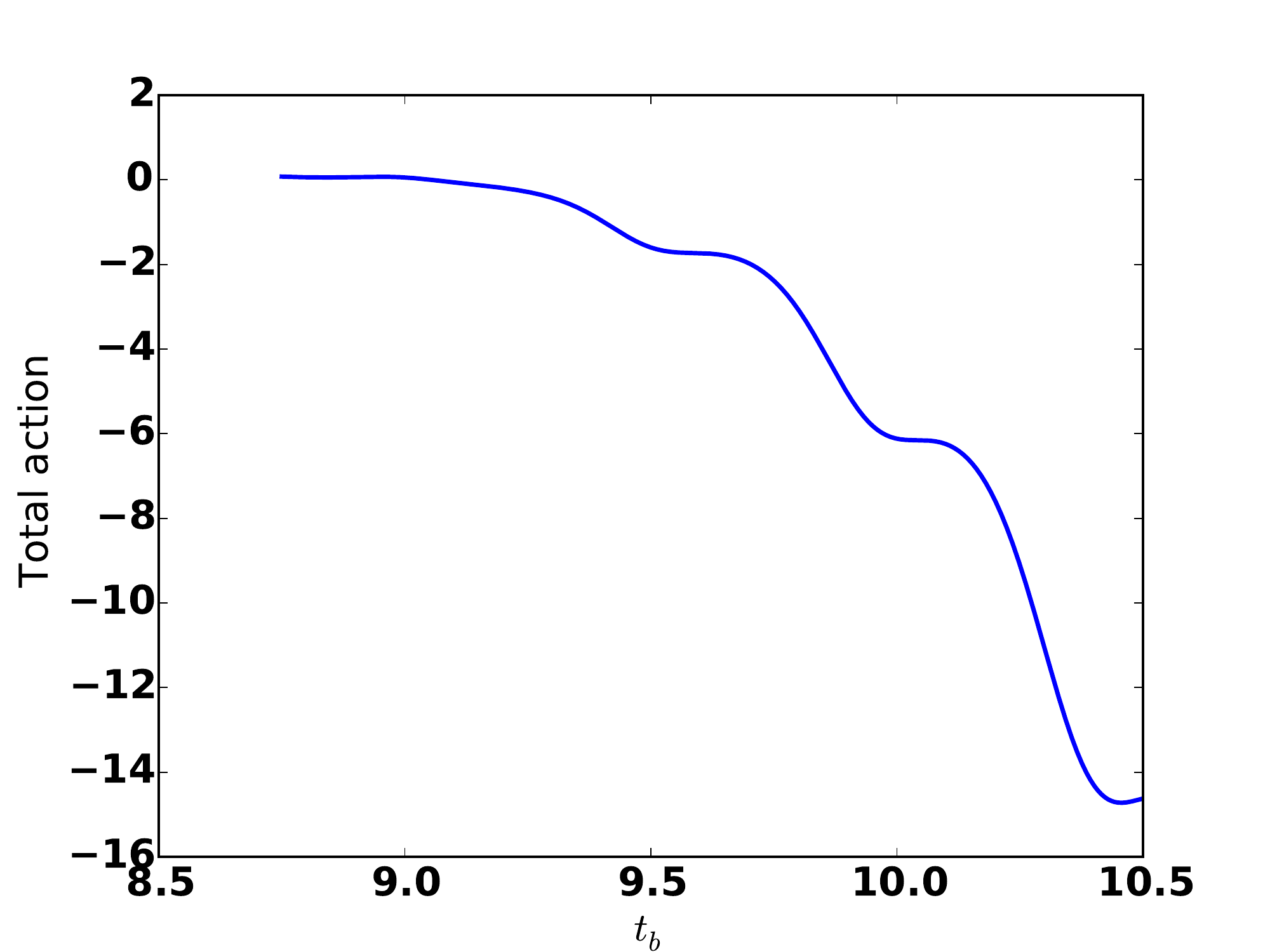}
 \caption{Total WdW action for the broad potential}
\end{subfigure}
 \caption{WdW actions for the steep and broad potentials.}\label{fig:HoloComp}
\end{figure}

We calculate numerically these actions, and plot the results in Figure~\ref{fig:HoloComp}.  In Figure 10 of \cite{HGST}, the authors find that, when the area of the leaf is flat as a function of $t_b$, the entanglement entropy decreases very slightly. We see in Figure~\ref{fig:HoloComp} that the gravitational WdW action decreases during exactly these periods of $t_b$--the complexity decreases when the entanglement entropy decreases. Moreover, the authors of \cite{HGST} found that the entanglement entropy decreased only very slightly. The decreases in the WdW gravitational action are bigger--they are some $O(1)$ fraction of the increases. 

For the ``steep" potential, the scalar action is much less than the gravitational action. However, for the broad potential this is not the case. Indeed, the scalar action is negative so that the total action is negative. Hence, the total action in this case appears to be unsuitable as a measure of circuit complexity. Therefore, in this case, the gravitational action (as opposed to the total action) behaves more like a complexity. Furthermore, we shall see below that the gravitational action qualitatively agrees with the expectations from the complexity-volume conjecture. This is in contrast to other settings, such as charged black holes in AdS, which require their total actions (gravitational plus Maxwell) to have similar behavior as complexity. Of course, there are many ways to define complexity in putative holographic duals to gravitational theories (depending on tolerance, gate set, etc.), so perhaps the different actions in the gravity theory correspond to different measures of complexity. Moreover, it is conceivable that the appropriate way to measure complexity in a holographic CFT is very different from the appropriate way to measure it in the holographic dual to FRW gravitational theories. This may explain why two different quantities behave like complexity in the two different settings. 

\begin{figure}[tbp]
\centering
\begin{subfigure}{0.45\textwidth}
 \includegraphics[width=\textwidth]{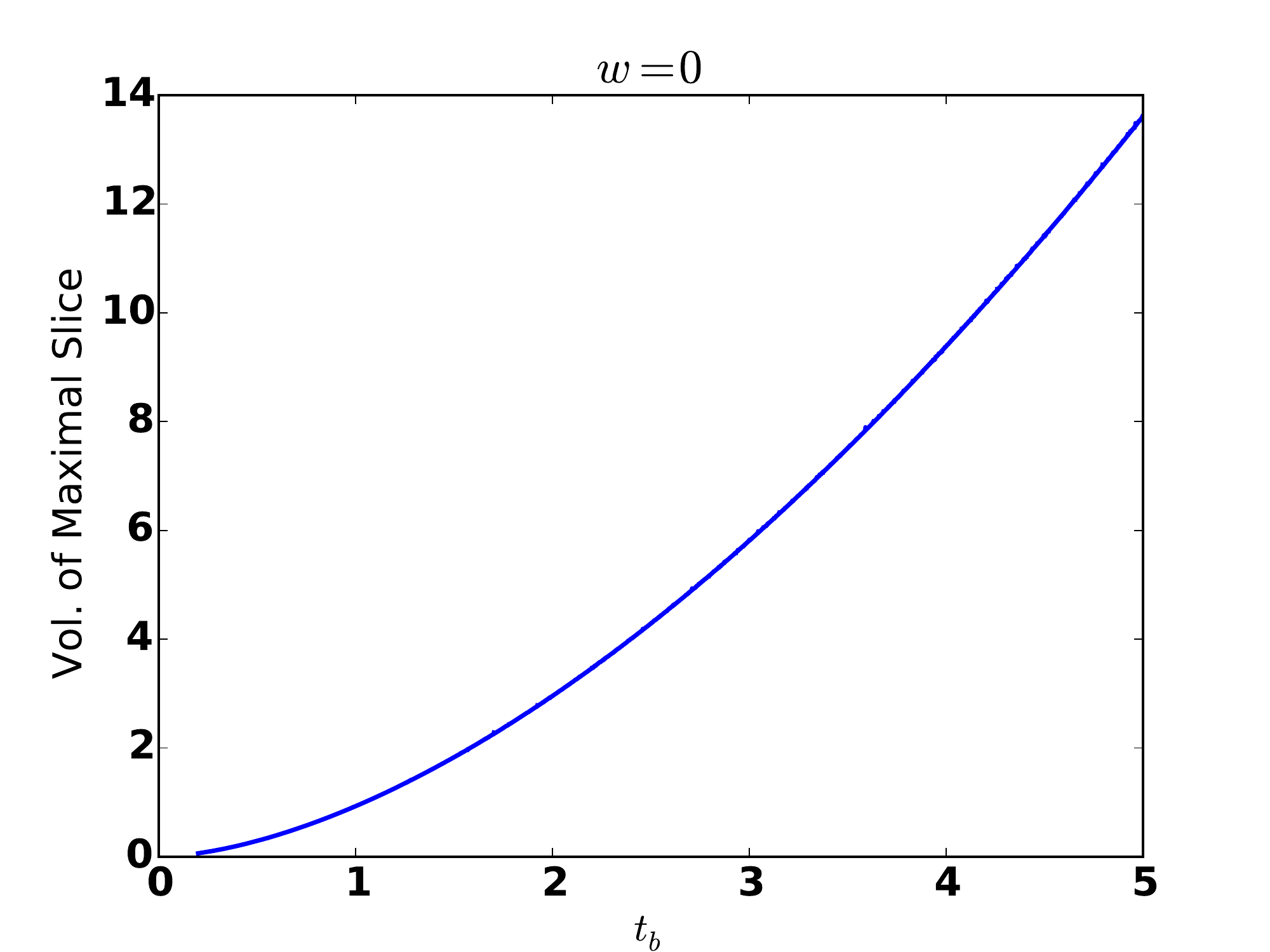}
 \caption{Maximal volume for a $w=0$ (matter) universe.}
\end{subfigure}
\begin{subfigure}{0.45\textwidth}
 \includegraphics[width=\textwidth]{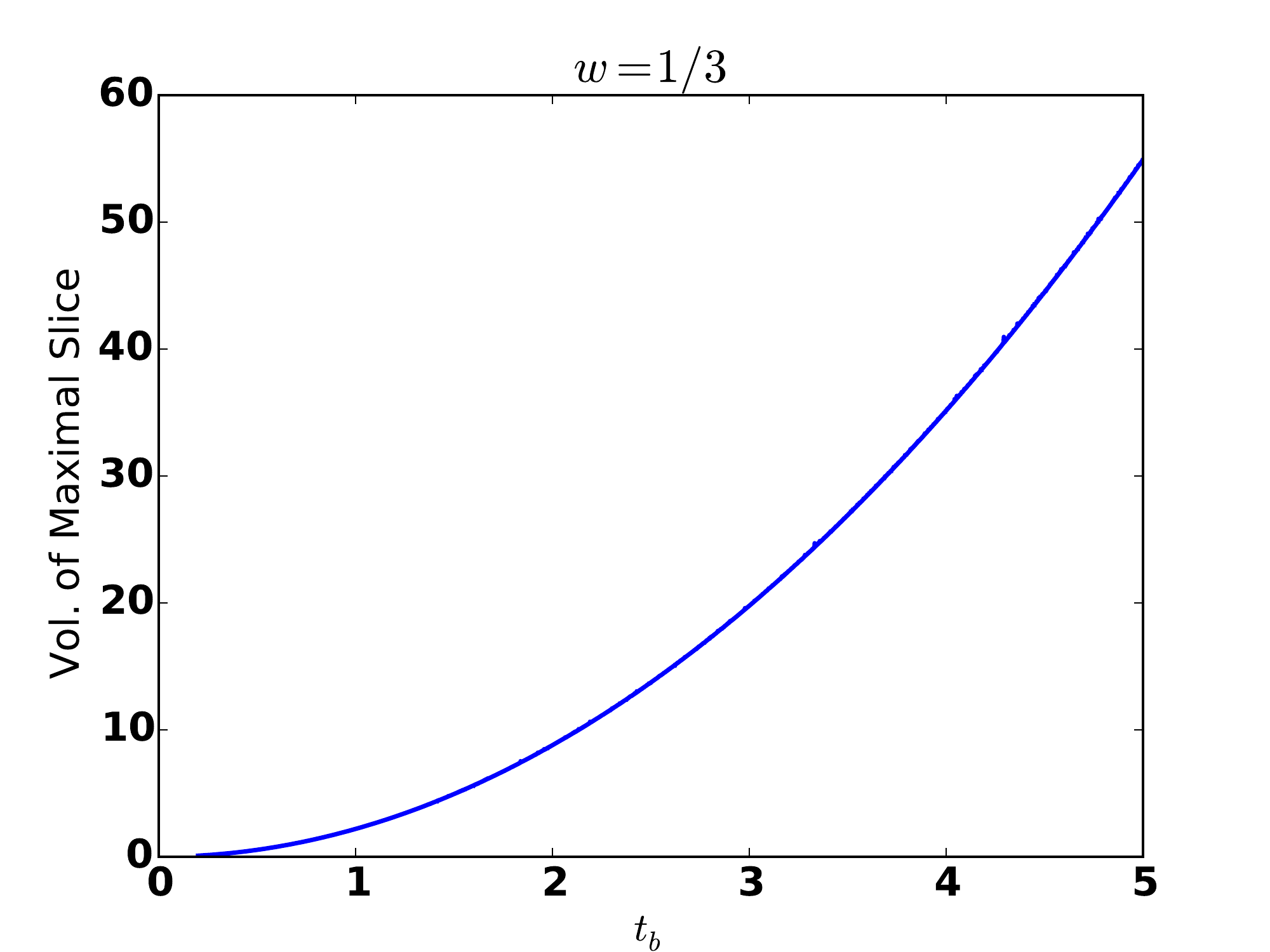}
 \caption{Maximal volume for a $w=1/3$ (matter) universe.}
\end{subfigure}
 
 \caption{Maximal volume vs boundary time for a single-component flat FRW Universe..}\label{fig:cv_single}
\end{figure}

\section{Complexity-Volume Conjecture for FRW Spacetimes}
We wish to analyze the complexity-volume conjecture. In the AdS/CFT context, this states that if a CFT state $\ket \psi$ at time $t$ is dual to some bulk geometry, then the complexity of $\ket \psi$ is given by the volume of the maximum-volume slice that is anchored at boundary time $t$ \cite{CompVol1,CompVol2}: 
$$\mathcal{C}( \ket \psi ) \sim \max V.$$
We examine the generalization of this conjecture to the FRW context by calculating the maximum volume slice inside of the holographic screen. We do this for an FRW universe dominated by one component, as well as one undergoing a transition. 

\subsection{Flat One-Component Universe}
In this case, we have one dominant matter component with equation of state 
$$P = w \rho.$$
The Friedmann equation gives 
$$a \propto t^{\frac{2}{3(1+w)}}.$$
The holographic screen, at time $t$ will be located at 
$$r=\frac{1}{\dot{a}(t)}.$$

We need to find the maximum-volume slice inside of $(t,r)$. This will be spherically-symmetric, and will have coordinates $r(\lambda), t(\lambda)$. The volume will be given by 
$$\text{Volume} =\int d \Omega_2 \int d \lambda r^2 \sqrt{- a(t)^2 \left (\frac{dt}{d \lambda} \right )^2+ \left (\frac{d r}{d \lambda} \right )^2 }.$$
Therefore, we need to extremize the functional 
$$I[t(\lambda), r(\lambda) ] =  \int d \lambda \sqrt{- r^4 a(t)^2 \left ( \frac{dt}{d \lambda} \right )^2+ a(t)^2 \left (\frac{d r}{d \lambda} \right )^2 },$$
subject to the initial conditions $r(0)=r,$ $t(0)=t$. This is done by solving the equations 
$$\frac{dr^2}{d \lambda^2} + \frac{2}{r}\left ( \frac{dr}{d \lambda} \right )^2+\frac{2}{r a(t)^2 }\left ( \frac{dt}{d \lambda} \right )^2 +  2 \frac{\dot{a}(t)}{a(t)} \frac{dr }{d \lambda} \frac{dt} {d \lambda}=0,$$
$$\frac{dt^2}{d \lambda^2} + a(t) \dot{a}(t) \left ( \frac{dr}{d \lambda} \right )^2 +   \frac{4}{r} \frac{dr }{d \lambda} \frac{dt} {d \lambda}=0.$$
We solve these equations numerically, and calculate the volumes of maximal slices for points on the holographic screen. We do this for matter (with $w=0$) and for radiation ($w=1/3$). The results are shown in Figure~\ref{fig:cv_single}. 

\begin{figure}[tbp]
\centering
\begin{subfigure}{0.45\textwidth}
 \includegraphics[width=\textwidth]{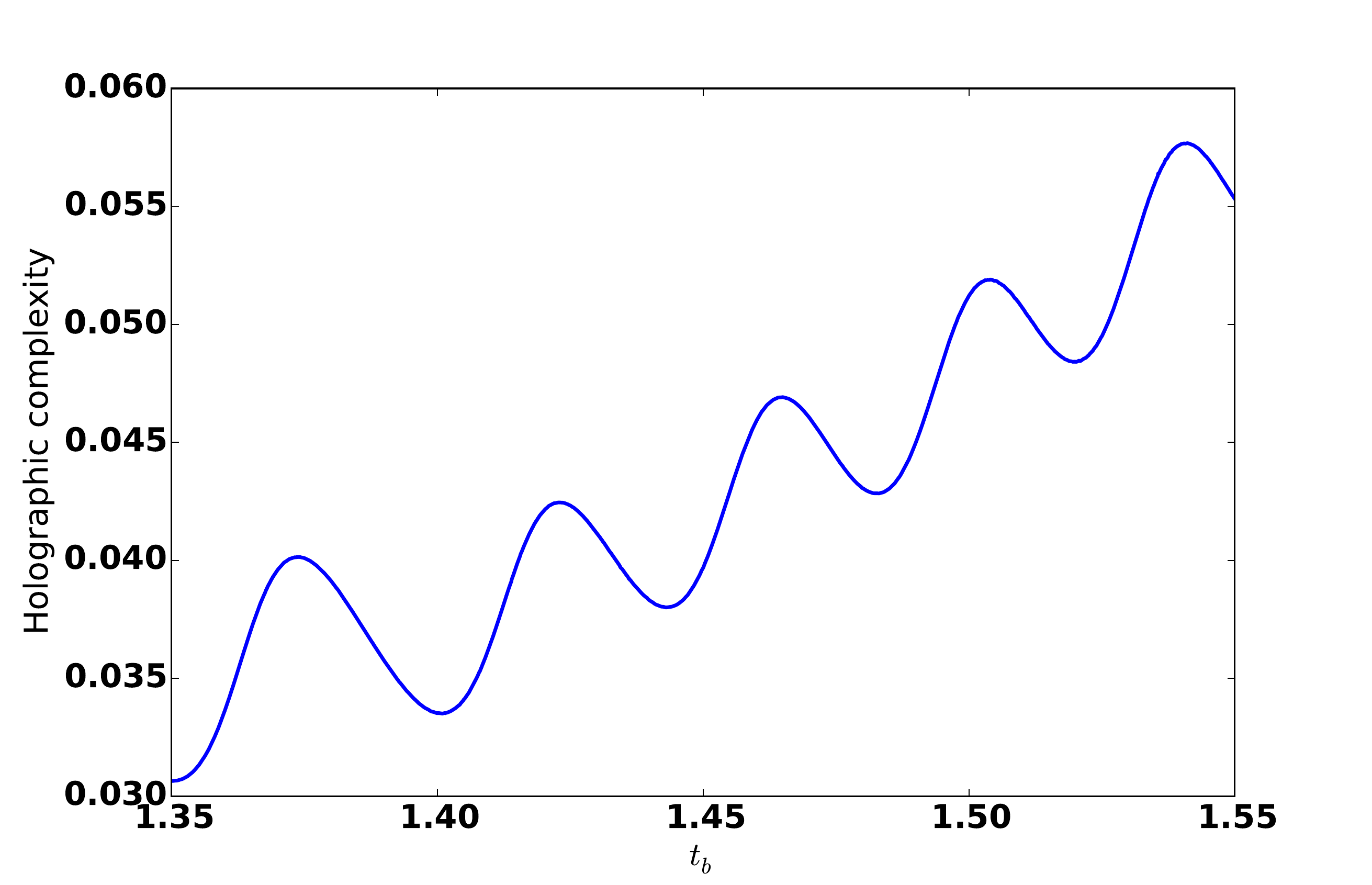}
 \caption{The action for the WdW patch for the steep potential}
\end{subfigure}
\begin{subfigure}{0.45\textwidth}
 \includegraphics[width=\textwidth]{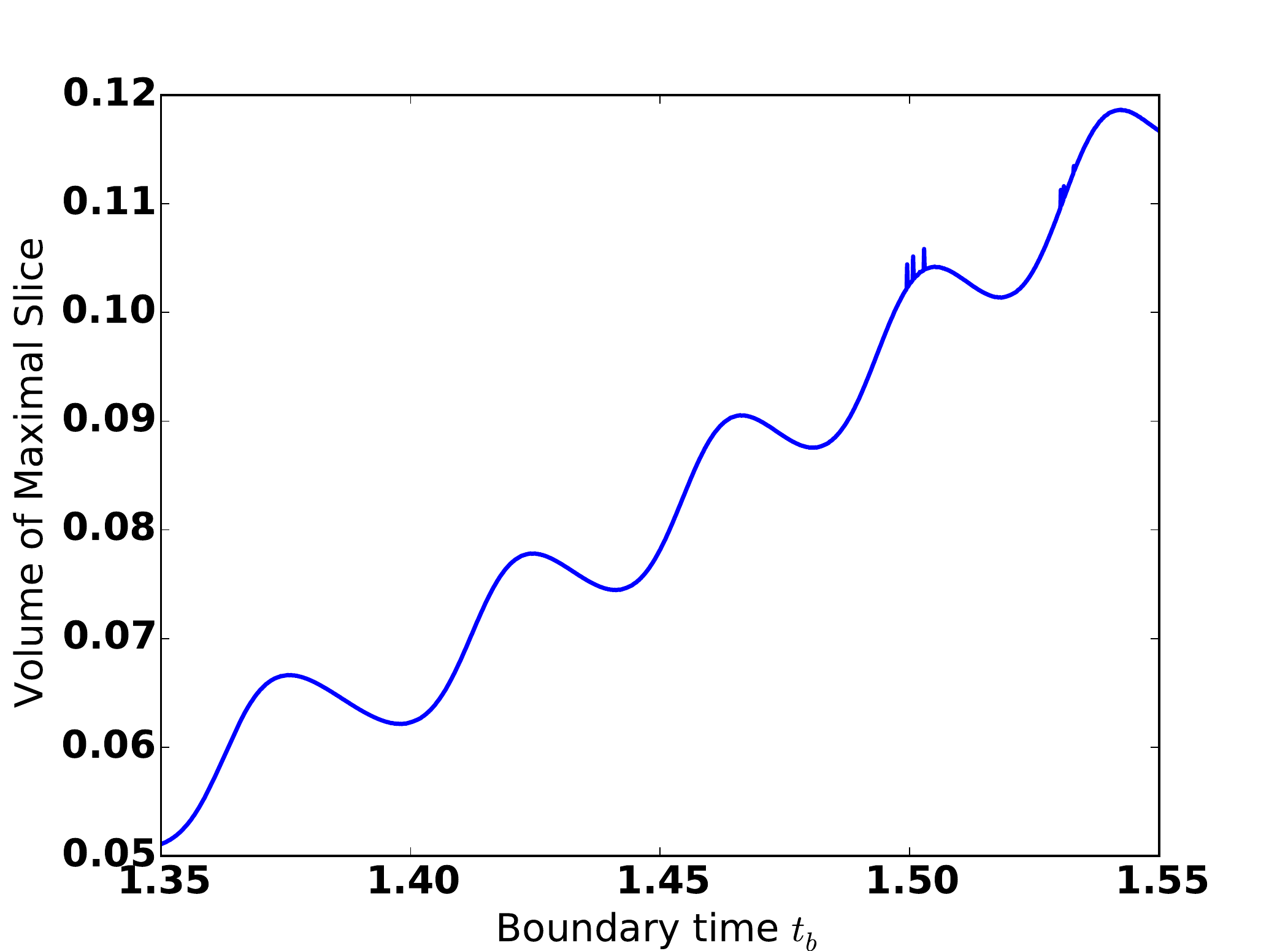}
 \caption{Maximum volume slice for the steep potential}
\end{subfigure}
 
 \caption{Holographic complexity for the steep and broad potentials.}\label{fig:HoloComp}
\end{figure}

\subsection{FRW Universe Undergoing a Transition}

We now consider the flat FRW universe that undergoes a transition, sourced by a scalar field. We previously saw that the bulk gravitational action of the WdW patch increases when the entanglement entropy increases, and decreases when the entanglement entropy decreases. We numerically solve for the scale factor for the previously-considered ``steep" potential, and then solve for the maximal slices. We show the results for the CV conjecture (as well as the results for the CA conjecture for comparison) in Figure~\ref{fig:HoloComp}. We see that the behavior is qualitatively very similar. The maximal volume increases and decreases essentially in the same time periods that the action of the WdW patch increases and decreases.

\section{The de Sitter Limit}

In this section, we analyze the de Sitter limit, which corresponds to $ w \rightarrow -1$. de Sitter space has natural thermodynamic variables associated to it, so it is interesting to see how this limit behaves with respect to those variables.

We have that the bulk WdW action is (for $w < -1/3$):

\begin{multline}\mathcal{I}^{WdW}_{bulk,gravity} =\frac{t_b^2}{4G_N} [2^{-\frac{6}{3 w+1}-2} (\frac{1}{3w+1}(9 (8^{\frac{2}{3
   w+1}+1} (3 w-1) \left(3 w \left(36 w^3+84
   w^2+w-86\right)-85\right)- \\ 16^{\frac{1}{3 w+1}} (3 w-5) (3 w+4) (3
   w+7) (1-3 w)^{\frac{2}{3 w+1}+3}+5 (w+1) (3 w+2) (1-3
   w)^{\frac{6}{3 w+1}+4}) (w+1))\\
   +9^{\frac{1}{3 w+1}}
   (2^{\frac{6}{3 w+1}+3} \left(-162 w^4-675 w^3-531 w^2+513
   w+515\right) (3 w+3)^{\frac{6 w}{3 w+1}}+ \\
   81 (w+1)^3 \left(5 (w+1)
   (3 w-1) (3 w+2) (3 w+3)^{\frac{4}{3 w+1}}+16^{\frac{1}{3 w+1}} (3
   w-5) (3 w+4) (3 w+7)\right)) (w+1)^{\frac{2}{3
   w+1}})] \\
   \cdot [3 (3 w-5) (3 w+2) (3 w+4) (3 w+7)]^{-1}.
\end{multline}
We take the de Sitter limit $w \rightarrow -1$ to find that 

$$ \lim_{w \rightarrow -1} \mathcal{I}^{WdW}_{bulk,gravity} = \frac{t_b^2}{4G_N} \frac{87}{1024}.$$
For a dS universe, we would perhaps expect that 
$$\frac{d \mathcal{C}}{dt} = TS,$$
where $T$ and $S$ are the standard thermodynamic values for the dS patch,
$$T= \frac{H}{2 \pi }, \text{			} S = \frac{\pi}{H^2},$$
where $H$ is the Hubble constant. Clearly, this does not agree with the limit $w \rightarrow -1$ of the FRW result; not even the scaling with $t$ matches. It is possible that this disagreement is to be expected, since the causal structure of the spacetime changes discontinuously at $w=-1$. This question merits further study. 

\section{Conclusions and Open Questions}

In this paper, we have examined the possible role of complexity in holographic theories of FRW spacetimes. We generalized the complexity-action and complexity-volume dualities from AdS to several FRW spacetimes. Specifically, for a flat FRW Universe with one component, we found that the WdW action grows as $t^2$, regardless of the matter content. This is to be contrasted with the behavior of the the WdW action for a black hole in AdS, where the complexity grows linearly with time. In the AdS case, the holographic theory is encoded in a set of degrees of freedom that remain fixed with time. However, in the FRW case, the holographic theory is encoded on the leaf $\sigma(t)$ which is growing in time, so one would expect the complexity to grow faster for the holographic dual to FRW spacetimes. For a FRW Universe sourced by a scalar field, undergoing a transition, the gravitational WdW action decreases during the time intervals when the entanglement entropy decreases. The fractional decreases in the gravitational action are much larger than the corresponding decreases in the entanglement entropy. This is consistent with the intuition that entanglement is computationally expensive. We then examined the complexity-volume conjecture, where we find similar qualitative behavior with the complexity-action results. While our work is speculative, we believe that the apparent consistency of the results are noteworthy. We close with some possible avenues of further study. 

First, it is interesting that, in the FRW cases, the quantity that behaves most like a complexity is the \textit{gravitational}, rather than the \textit{total} action.  In other settings, for example, charged AdS black holes, it is the total (gravity plus Maxwell) action that seems to be dual to complexity \cite{Brown}.  In the FRW setting, the total action is negative, which is of course not a sensible result for complexity. Complexity in the putative holographic theory is, of course, only defined up to choices in parameters like the tolerance, gate set, etc. It is conceivable that the different WdW actions correspond to different definitions of complexity. Furthermore, since a holographic CFT and the holographic dual to the FRW universe are very different theories (see, for example, \cite{Unentanglement}), it is plausible that they require different measures of complexity. The precise holographic dictionary between actions and volumes and complexities clearly merit further research. Our results provide more explicit tests of holographic complexity conjectures, which may lead to clarification on this dictionary. 

It is also noteworthy that in the case of an FRW universe with one component, the scaling of the complexity with time is $t^2$, independent of the value of $w.$ This is reminiscent of the butterfly velocities in holographic theories of FRW spacetimes \cite{Butterfly}. It would be interesting to explore the relationship between butterfly velocities and complexity, and to see if there is some common explanation for the $w$-independence of the scaling of these two quantities. 

It would also be interesting to try to better understand the de Sitter (i.e., $w \rightarrow -1$) limit. In particular, the holographic screen degenerates to the past/future infinity of dS space. It is possible that a better understanding of the FRW Universe and its $w \rightarrow -1$ limit will lead to an improved understanding of quantum gravity in dS. 

Finally, it is has been argued that the holographic theory of FRW spacetimes fall into one of two structures, which have been termed the ``Russian nesting doll" structure, and the ``spacetime equals entanglement" structure \cite{HGST}. It would be interesting to see if the results obtained in this paper could be useful in obtaining a better understanding of the way the bulk FRW spacetimes are encoded in their boundary theories, and the structure of their Hilbert spaces. 

\acknowledgments

I am grateful to Matt Headrick and Yasunori Nomura for comments on the manuscript, as well as many fruitful discussions. I would also like to thank Sam Leutheusser, Leonard Susskind, and Brian Swingle for useful conversations and correspondences.

\appendix
\section{Boundary and Corner Terms in the Gravitational Action}
\label{app:A}
Consider a spacetime region $W$. Its gravitational action will consist of the bulk Einstein-Hilbert and boundary Gibbons-Hawking-York term, as well as null boundary terms and corner terms. This prescription was first worked out in \cite{Corner}. Suppose the boundary $\partial W$ consists of several smooth pieces that are connected at ``corners." Consider one such smooth, null component $N$, with corners $B_1$ and $B_2$. The boundary and corner terms are:

\begin{multline}
 \mathcal{I}_N = \frac{sgn(N)}{8 \pi G_N} \int_N d \lambda d \theta \sqrt{\gamma} \kappa + \frac{sgn(N)}{8 \pi G_N} \int_N d \lambda d \theta \sqrt{\gamma} \Theta \log ( l_c \abs{\Theta}) \\ + \frac{1}{8 \pi G_N} \int_{B_1} d \theta \sqrt{\gamma} a_1 +\frac{1}{8 \pi G_N} \int_{B_2} d \theta \sqrt{\gamma} a_2,
\end{multline}
where $sgn(N)$ is +1 if $N$ is to the future of $W$, and -1 if it is to the past of $W$. The quantities $a_1, a_2$ depend on the precise nature of the joint. In the case where we have a joint between two null surfaces, with normal vectors $k$ and $m$, the joint will have
$$a = \pm \log \abs {\frac{ k \cdot m}{2}}.$$
The sign is determined is follows. If $W \subset J^+(N)$, and the corner is at the past end of $N$, or if $W \subset J^-(N)$, and the corner is at the future end of $N$, then the sign is positive. In all other cases, it is negative. 

As was first demonstrated in \cite{Corner}, the above action is independent of the parameterization of the null generators of $N$. We demonstrate that here. Suppose, for definiteness, that $W$ lies to the past of $N$. Let $B_2$ be at the future end of $N$, and let $B_1$ be at its past end. The boundary and corner terms associated to $N$ are given by
\begin{multline}
 \mathcal{A}_N = \frac{-1}{8 \pi G_N} \int_N d \lambda d \theta \sqrt{\gamma} \kappa - \frac{1}{8 \pi G_N} \int_N d \lambda d \theta \sqrt{\gamma} \Theta \log ( l_c \abs{\Theta}) \\ - \frac{1}{8 \pi G_N} \int_{B_1} d \theta \sqrt{ \gamma } A_1 + \frac{1}{8 \pi G_N} \int_{B_2} d \theta \sqrt{ \gamma } A_2,
\end{multline}  
where $A_i$ denotes the positive sign of $a_i.$

We now show that $\mathcal{I}_N$ is independent of the geodesic parameterization. Suppose that we parameterize the null geodesics by a new parameter, $\bar \lambda ( \lambda , \theta )$, and define $e^ {- \beta} \equiv d \bar \lambda / d \lambda.$ Clearly, we have 
$$ \bar k = \frac{\partial \lambda }{\partial \bar \lambda} k = e^\beta k,$$
and that 
$$\bar \Theta = \frac{1}{\sqrt{\gamma}} \frac{ \partial \sqrt{\gamma}}{\partial \bar \lambda} = e^ \beta \Theta.$$
We define the vector $N^\alpha$ to be the null vector such that $k^\alpha N_\alpha = -1$ so that 
$$ \kappa = - N_\nu k^\mu \nabla_\mu k^\nu.$$
Of course, $\bar N = e^{- \beta} N$ (since $\bar{k} \cdot \bar{N} = -1$), which means that 
$$\bar \kappa = - N_\nu k^\mu \nabla_\mu (e^\beta k^\nu) =e^\beta \kappa -N_\nu k^\nu k^\mu \nabla_\mu e^\beta = e^{\beta} ( \kappa + \partial_\lambda \beta) .$$

Thus, since $\bar A = A + \beta,$ we obtain
\begin{multline}
\bar{\mathcal{I}}_N = -\frac{1}{ 8 \pi G_N} \int_N d \lambda d \theta \sqrt{\gamma} (\kappa + \partial_\lambda \beta) - \frac{1}{8 \pi G_N} \int_N d \lambda d \theta \sqrt{\gamma} \Theta \log ( l_c \abs{\Theta}) - \frac{1}{8 \pi G_N} \int_N d \lambda d \theta \sqrt{\gamma} \Theta \beta \\
- \frac{1}{8 \pi G_N} \int_{B_1} d \theta \sqrt{ \gamma } A_1 +\frac{1}{8 \pi G_N} \int_{B_2} d \theta \sqrt{ \gamma } A_2 \\ -\frac{1}{8 \pi G_N} \int_{B_1} d \theta \sqrt{ \gamma } \beta +\frac{1}{8 \pi G_N} \int_{B_2} d \theta \sqrt{ \gamma } \beta\\
= -\frac{1}{ 8 \pi G_N} \int_N d \lambda d \theta \sqrt{\gamma} \kappa  + \frac{1}{ 8 \pi G_N} \int_N d \lambda d \theta \beta \partial_\lambda  \sqrt{\gamma}  - \frac{1}{8 \pi G_N} \int_N d \lambda d \theta \sqrt{\gamma} \Theta \log ( l_c \abs{\Theta}) - \frac{1}{8 \pi G_N} \int_N d \lambda d \theta \sqrt{\gamma} \Theta \beta \\
- \frac{1}{8 \pi G_N} \int_{B_1} d \theta \sqrt{ \gamma } A_1 +\frac{1}{8 \pi G_N} \int_{B_2} d \theta \sqrt{ \gamma } A_2 \\
=-\frac{1}{ 8 \pi G_N} \int_N d \lambda d \theta \sqrt{\gamma} \kappa   - \frac{1}{8 \pi G_N} \int_N d \lambda d \theta \sqrt{\gamma} \Theta \log ( l_c \abs{\Theta})  \\
- \frac{1}{8 \pi G_N} \int_{B_1} d \theta \sqrt{ \gamma } A_1 +\frac{1}{8 \pi G_N} \int_{B_2} d \theta \sqrt{ \gamma } A_2 =\mathcal{I}_N.
\end{multline}
Hence, we see that $\mathcal{I}_N$ is invariant under reparameterization.

\end{document}